\documentclass[preprint,3p]{elsarticle}
\usepackage{graphics}
\usepackage{dcolumn}
\usepackage{amssymb}
\usepackage{amsthm}
\usepackage[mathlines]{lineno}
\usepackage{graphicx}
\usepackage{units}
\usepackage{url}
\usepackage{amsmath}
\usepackage{amsfonts}
\usepackage{bm}
\usepackage{textcomp}
\usepackage{subfigure}
\usepackage{multicol}
\usepackage{verbatim}
\usepackage[colorlinks,linkcolor=blue,citecolor=blue]{hyperref}
\usepackage{rotating}
\usepackage{lscape}
\usepackage{overpic}

\newcommand{\jpsi}{J/\psi}
\newcommand{\psip}{\psi(2S)}
\newcommand{\elp}{e^{+}}
\newcommand{\elm}{e^{-}}
\newcommand{\mup}{\mu^{+}}
\newcommand{\mum}{\mu^{-}}
\newcommand{\pip}{\pi^{+}}
\newcommand{\pim}{\pi^{-}}
\newcommand{\piz}{\pi^{0}}

\newcommand{\rhoz}{\rho^{0}}
\newcommand{\prp}{p}
\newcommand{\prm}{\bar{p}}
\newcommand{\kp}{K^{+}}
\newcommand{\km}{K^{-}}

\newcommand{\chisq}[1]{\chi^{2}_{#1}}

\newcommand{\Xf}{X_{f}}

\newcommand{\fivepi}{2(\pi^{+}\pi^{-})\pi^{0}}

\journal{Physics Letters B}

\begin{document}

\begin{frontmatter}

%\begin{lstlisting}
%\normalsize

%\parskip=5pt plus 1pt minus 1pt

%\linenumbers

\title{{\bf
\boldmath Measurement of the phase between strong and electromagnetic amplitudes of $\jpsi$ decays}}

\author{
\begin{small}
\begin{center}
M.~Ablikim$^{1}$, M.~N.~Achasov$^{9,d}$, S. ~Ahmed$^{14}$, M.~Albrecht$^{4}$, A.~Amoroso$^{53A,53C}$, F.~F.~An$^{1}$, Q.~An$^{50,40}$, Y.~Bai$^{39}$, O.~Bakina$^{24}$, R.~Baldini Ferroli$^{20A}$
% ...
\\
\vspace{0.2cm}
(BESIII Collaboration)\\
\vspace{0.2cm} {\it
$^{1}$ Institute of High Energy Physics, Beijing 100049, People's Republic of China\\
% ...
$^{4}$ Bochum Ruhr-University, D-44780 Bochum, Germany\\
% ...
$^{9}$ G.I. Budker Institute of Nuclear Physics SB RAS (BINP), Novosibirsk 630090, Russia\\
% ...
$^{14}$ Helmholtz Institute Mainz, Johann-Joachim-Becher-Weg 45, D-55099 Mainz, Germany\\
% ...
$^{20}$ (A)INFN Laboratori Nazionali di Frascati, I-00044, Frascati, Italy; (B)INFN and University of Perugia, I-06100, Perugia, Italy\\
% ...
$^{24}$ Joint Institute for Nuclear Research, 141980 Dubna, Moscow region, Russia\\
% ...
$^{39}$ Southeast University, Nanjing 211100, People's Republic of China\\
$^{40}$ State Key Laboratory of Particle Detection and Electronics, Beijing 100049, Hefei 230026, People's Republic of China\\
% ...
$^{50}$ University of Science and Technology of China, Hefei 230026, People's Republic of China\\
% ...
$^{53}$ (A)University of Turin, I-10125, Turin, Italy; (B)University of Eastern Piedmont, I-15121, Alessandria, Italy; (C)INFN, I-10125, Turin, Italy\\
% ...
$^{d}$ Also at the Novosibirsk State University, Novosibirsk, 630090, Russia\\
% ...
}\end{center}
\vspace{0.4cm}
\end{small}
}
%\affiliation{}

\begin{abstract}
Using 16 energy points of $\elp\elm$ annihilation data collected
in the vicinity of the $\jpsi$ resonance with the BESIII detector and with
a total integrated luminosity of around 100 pb$^{-1}$,
we study the relative phase between the strong and electromagnetic amplitudes of $\jpsi$ decays.
The relative phase between $\jpsi$ electromagnetic decay and
the continuum process ($\elp\elm$ annihilation without the $\jpsi$ resonance) is
confirmed to be zero by studying the cross section lineshape of $\mup\mum$ production.
The relative phase between $\jpsi$ strong and electromagnetic decays
is then measured to be $(84.9\pm3.6)^\circ$ or $(-84.7\pm3.1)^\circ$
for the $\fivepi$ final state by investigating the interference pattern between the
$\jpsi$ decay and the continuum process.
This is the first measurement of the relative phase between $\jpsi$ strong and
electromagnetic decays into a multihadron final state
using the lineshape of the production cross section.
We also study the production lineshape of the multihadron final state
$\eta\pip\pim$ with $\eta\to\pip\pim\piz$, which provides additional information
about the phase between the $\jpsi$ electromagnetic decay amplitude and the continuum process.
Additionally, the branching fraction of $\jpsi\to\fivepi$ is measured to be
$(4.73\pm0.44)\%$ or $(4.85\pm0.45)\%$, and the branching fraction of
$\jpsi\to\eta\pip\pim$ is measured to be $(3.78\pm0.68)\times10^{-4}$. Both of them are
consistent with the world average values. The quoted uncertainties include both statistical and
systematic uncertainties, which are mainly caused by the low statistics.

\end{abstract}

\end{frontmatter}
%\pacs{25.75.Gz, 14.40.Df, 12.38.Mh}

%\linenumbers
\section{INTRODUCTION}

\begin{figure}[htbp]
\begin{center}
    \includegraphics[angle=0,width=4.5cm, height=1.8cm]{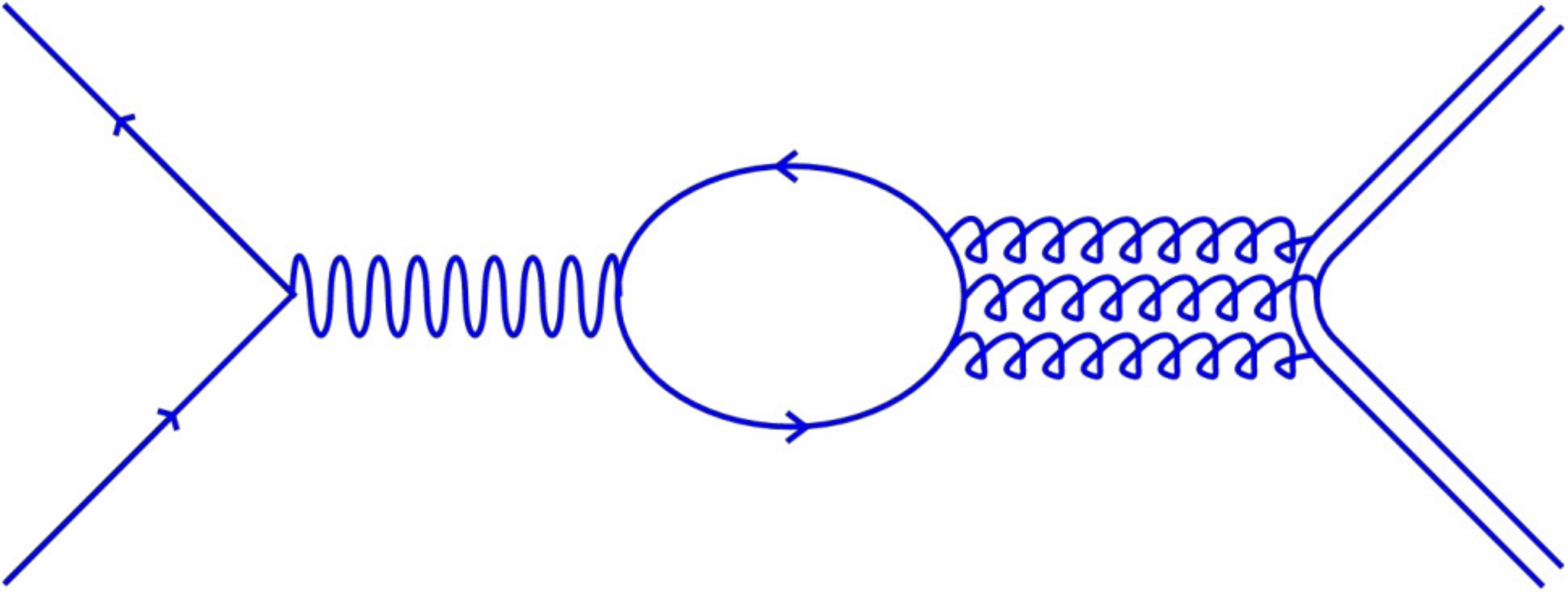}
    \put(-80, 45){(a)} \put(-50,40){$A_{g}$}
    \hskip 0.5cm
    \includegraphics[angle=0,width=4.5cm, height=1.8cm]{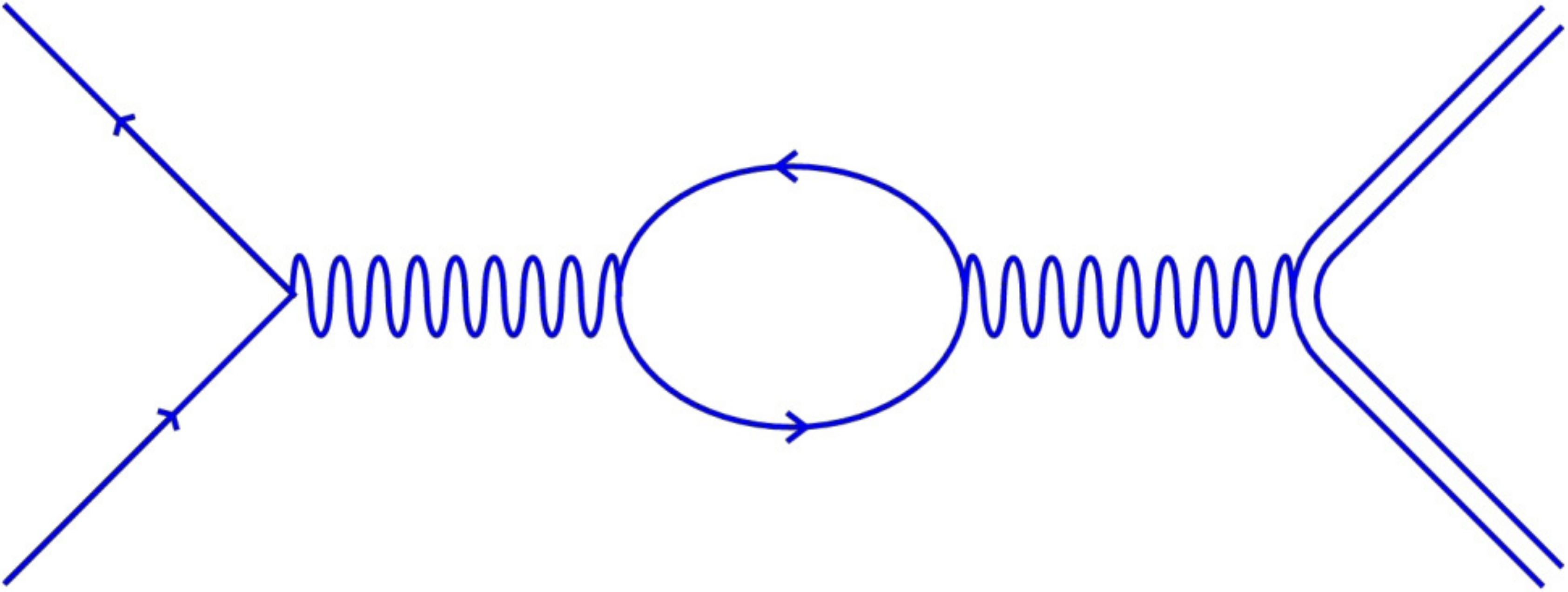}
    \put(-80, 45){(b)} \put(-50,40){$A_{\gamma}$}
    \hskip 5mm
    \includegraphics[angle=0,width=4.5cm, height=1.8cm]{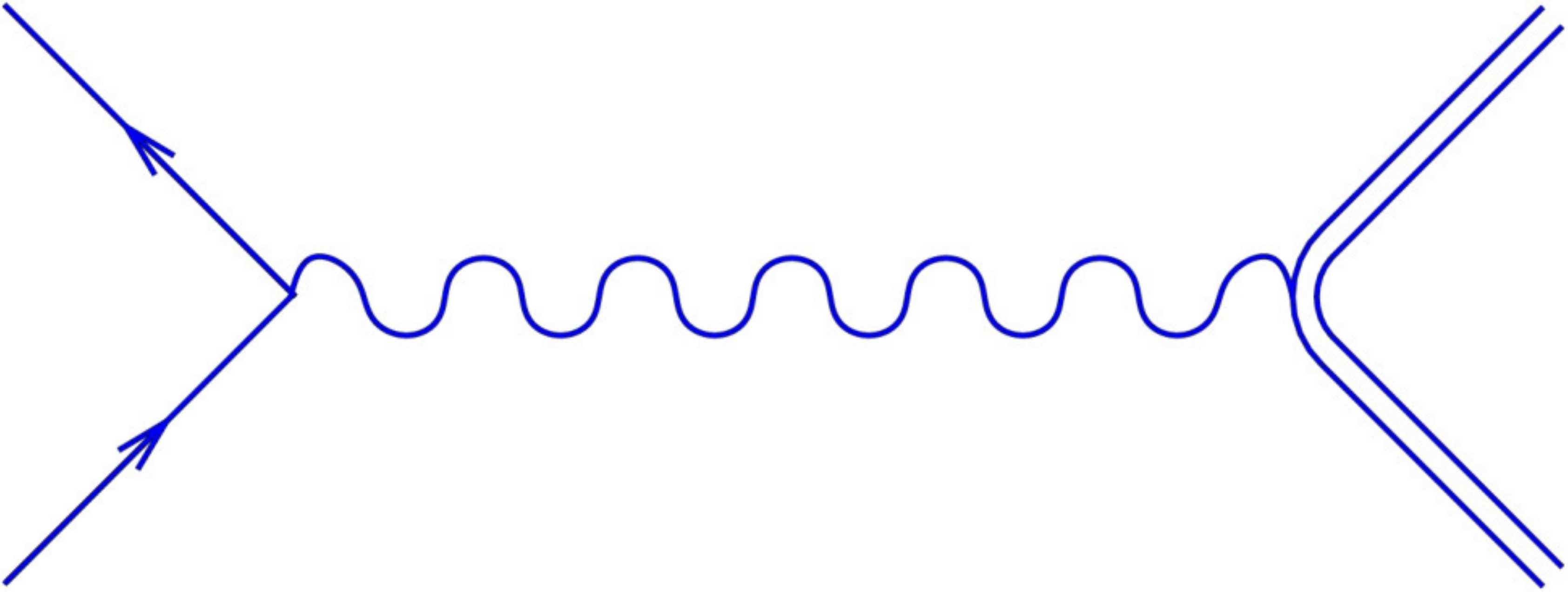}
    \put(-80, 45){(c)} \put(-50,40){$A_{\rm cont}$}
 \caption{The Feynman diagrams for the process $\elp\elm\to$~hadrons: (a) $\jpsi$ strong decay {\it via} gluons,
 (b) $\jpsi$ EM decay {\it via} one virtual photon, (c) the continuum decay {\it via} a virtual photon.}
 \label{fig_diagrams}
\end{center}
\end{figure}
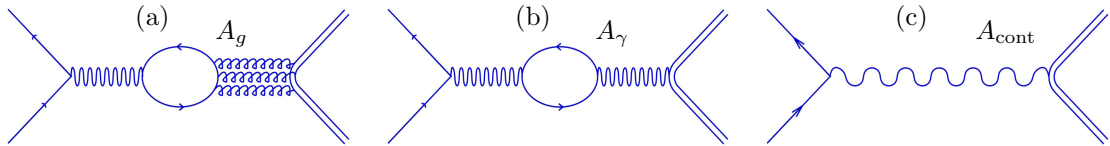

The relative phase between the strong and electromagnetic (EM) amplitudes of quarkonium
decays is a basic parameter that provides insight into the dynamics of quarkonium decays.
As shown in Fig.~\ref{fig_diagrams}, in the vicinity of the $\jpsi$,
the annihilation of $\elp\elm$ into a hadronic final state proceeds through three processes:
strong decay of the $J/\psi$~(mediated by gluons), EM decay of the $J/\psi$~(mediated by a virtual photon),
and the continuum process~(without a $\jpsi$ intermediate state and mediated by a virtual photon).
For leptonic final states, on the other hand, the strong decay is absent.
In perturbative quantum chromodynamics, the relative phase~($\Phi_{g,\gamma}$)
between the charmonium strong decay amplitude~($A_{g}$) and the EM amplitude~($A_\gamma$)
is predicted to be $0^{\circ}$ or $180^{\circ}$~\cite{brodsky,ref_pqcd} at lowest order.

In contrast to this prediction, model-dependent analyses using SU(3) flavor symmetry suggest
that $\Phi_{g,\gamma}$ is $90^{\circ}$ for $\jpsi$ two-body decays into meson pairs
with quantum numbers~($J^{\rm P}$) of $1^{-}0^{-}$~\cite{jousset,coffman},
$0^{-}0^{-}$~\cite{suzuki1,lopez,kopke}, $1^{-}1^{-}$~\cite{kopke}, and $1^{+}0^{-}$~\cite{suzuki2},
and for $\jpsi$ decays into $N\bar{N}$ baryon pairs~\cite{rinaldo1,zhuk}.
Similar analyses suggest $\psip$ decays to pairs of pseudoscalar mesons also have a phase
$\Phi_{g,\gamma}$ around $90^{\circ}$, but $\psip$ decays to pairs of mesons with $1^{-}0^{-}$
and $1^{+}0^{-}$ have a different value of $\Phi_{g,\gamma}$~\cite{suzuki2,chen}.

Several theoretical ideas regarding the origin and implications of $\Phi_{g,\gamma}$ have been proposed.
Based on unsubtracted dispersion relations and asymptotic freedom,
the {Okubo-Zweig-Iizuka-rule-violating} amplitude with respect to the virtual photon
contribution is predominately imaginary~\cite{ref_Fukugita}.
An orthogonal phase in $\jpsi$ decays is also expected if any vector quarkonium
is assumed to be coupled to a vector glueball~\cite{hou,nambu,rinaldo2}.
Furthermore, it has been advocated~\cite{suzuki2, rinaldo2} that different phases for
the $\jpsi$ and $\psip$ decay, namely $\sim 90^{\circ}$ and $\sim 180^{\circ}$, respectively,
can explain the long-standing $\rho\pi$ puzzle of charmonium physics.
{However, there is no simple explanation that the phase should be $90^{\circ}$.}

An independent approach for measuring the relative phases of the diagrams in Fig.~\ref{fig_diagrams}
consists of extracting the interference pattern of the $\elp\elm$ reaction cross section as a
function of the center-of-mass (CM) energy ($W$) in the vicinity of a resonance.
The Born cross section of a pure EM process can be written as
\begin{equation}
\sigma^{0}(W) {\propto} |A_{\gamma}(W)e^{i\Phi_{\gamma,\rm cont}}+A_{\rm cont}(W)|^{2}.\notag
\label{eq_sum1}
\end{equation}
The relative phase~($\Phi_{\gamma,\rm cont}$) between the $\jpsi$ EM amplitude~($A_{\gamma}$) and the continuum amplitude~($A_{\rm cont}$) has previously been assumed to be zero degrees and this assumption has been shown to be consistent with the observed interference pattern in $\jpsi$ decays to lepton pairs~\cite{ref_slc,ref_bes,ref_kder,BW}.
The full Born cross section for processes including the strong and EM amplitudes can be written as
\begin{equation}
\sigma^{0}(W) {\propto} |[A_{g}(W)e^{i\Phi_{g,\gamma}}+A_{\gamma}(W)]e^{i\Phi_{\gamma,\rm cont}}+A_{\rm cont}(W)|^{2}. \notag
\label{eq_sum}
\end{equation}
If we take the phase $\Phi_{\gamma,\rm cont}$ to be zero, as measurements suggest, the Born cross section is simplified to be:
\begin{equation}
\sigma^{0}(W) {\propto} |A_{g}(W)e^{i\Phi_{g,\rm EM}}+A_{\gamma}(W)+A_{\rm cont}(W)|^{2},\notag
\label{eq_sum2}
\end{equation}
where $\Phi_{g,\rm EM}$ is the phase between the strong and the full EM amplitudes.

It is argued that the relative phases $\Phi_{\gamma,\rm cont}$ and $\Phi_{g,\rm EM}$ are universal in all exclusive decay modes~\cite{gerard}.
In this Letter, we first analyze the process $\elp\elm\to\mup\mum$ and confirm the phase \textcolor{black}{$\Phi_{\gamma,\rm cont}$} is consistent with zero.
We also use this process to extract the CM energy spread and the overall energy scale,
which are essential accelerator parameters that are used as input for the other analyses.
Then, we measure the phase $\Phi_{g,\rm EM}$ by analyzing the process $\elp\elm\to\fivepi$ (abbreviated as $5\pi$).
We chose this process because it both has a large branching fraction in $\jpsi$ decays and has a sizable cross section of the continuum decay.
We also study the process $\elp\elm\to\eta\pip\pim$ with $\eta$ decaying into $\pip\pim\piz$.
Since it proceeds largely through {$\eta\rhoz$, which is an EM process due to G-parity conservation,} this process
is used to gain further information about $\Phi_{\gamma,\rm cont}$.
This is the first measurement of the phases $\Phi_{g,\rm EM}$ and $\Phi_{\gamma,\rm cont}$ in the
interference pattern of the cross section lineshape in the vicinity of the $\jpsi$ and
the first time using {multihadron} final states.

This letter is organized as follows: in Section~\ref{detector}, the BESIII detector and the data sets being used are described.
In Section~\ref{analysis}, the event selection, the efficiency, the observed cross section and the systematic uncertainties
of $\elp\elm\to\mup\mum$, $5\pi$ and $\eta\pip\pim$ are described.
In Section~\ref{results}, the fit to the cross section lineshapes of $\elp\elm\to\mup\mum$, $5\pi$ and $\eta\pip\pim$ as well as the results
are reported.
The results are summarized in Section~\ref{summary}.

\section{BESIII EXPERIMENT AND DATA SETS} \label{detector}

The BEPCII is a double-ring $\elp\elm$ collider running at CM energies between $2.0-4.6$~GeV
and it has reached its design luminosity of $1.0\times10^{33}$~cm$^{-2}$s$^{-1}$
at a CM energy of $3770$~MeV. The cylindrical BESIII detector has an effective geometrical
acceptance of 93\% of $4\pi$ solid angle and it is divided into a barrel section and two endcaps.
It consists of a small-cell, helium-based {multilayer} drift chamber (MDC),
a plastic scintillator time-of-flight system (TOF), a CsI(TI) (Thallium doped Cesium Iodide)
crystal electromagnetic calorimeter (EMC) and a muon system containing resistive plate chambers in
the iron return yoke of a 1 Tesla (0.9 Tesla for data sets used in this letter) superconducting solenoid.
The momentum resolution for charged tracks is 0.5\% for 1 GeV$/c$ momentum tracks.
The time resolution in the barrel (endcaps) is 80 ps (110 ps).
The photon energy resolution at 1 GeV is 2.5\% (5\%) in the barrel (endcaps) of the EMC.
Further details about the BESIII detector are described in Ref.~\cite{detector}.

This analysis uses data samples collected in 2012 at 16 different CM energies with a total integrated luminosity of about 100~pb$^{-1}${~\cite{zhang_lumi}}.
The CM energies, $W_i$, and the integrated luminosities, $\mathcal{L}_{i}$, of each data sample are summarized in Table~\ref{table_data set}.
The CM energies are measured by the Beam Energy Measurement System (BEMS),
in which photons from a CO$_2$ laser are Compton back-scattered off the electron beam and detected by a high-purity Germanium detector~\cite{bems}.
The integrated luminosities are determined using two-gamma events~\cite{zhang_lumi}.

\begin{table}[htp]
\begin{center}
\caption{\small The CM energy ($W_{i}$) and the integrated luminosity ($\mathcal{L}_{i}$) for each data point.
The uncertainty of $W_{i}$ is from the BEMS measurement, and the uncertainty of $\mathcal{L}_{i}$ is the statistical and systematic uncertainties added in quadrature{~\cite{zhang_lumi}}.}
\label{table_data set}
\begin{tabular}{{c}{c}{c}{c}{c}{c}{c}{c}{c}ccc} \hline
No.  &$W_{i}$ (MeV)      &$\mathcal{L}_{i}$~(pb$^{-1}$)  \\\hline
1   &$3050.21\pm0.03$    &$14.92\pm0.16$  \\
2   &$3059.26\pm0.03$    &$15.06\pm0.16$  \\
3   &$3080.20\pm0.02$    &$17.39\pm0.19$  \\
4   &$3083.06\pm0.04$    &$\phantom{3}4.77 \pm0.06$  \\
5   &$3089.42\pm0.02$    &$15.56\pm0.17$  \\
6   &$3092.32\pm0.03$    &$14.91\pm0.16$  \\
7   &$3095.26\pm0.08$    &$\phantom{3}2.14 \pm0.03$  \\
8   &$3095.99\pm0.08$    &$\phantom{3}1.82 \pm0.02$  \\
9   &$3096.39\pm0.08$    &$\phantom{3}2.14 \pm0.03$  \\
10  &$3097.78\pm0.08$    &$\phantom{3}2.07 \pm0.03$  \\
11  &$3098.90\pm0.08$    &$\phantom{3}2.20 \pm0.03$  \\
12  &$3099.61\pm0.09$    &$\phantom{3}0.76 \pm0.01$  \\
13  &$3101.92\pm0.11$    &$\phantom{3}1.61 \pm0.02$  \\
14  &$3106.14\pm0.09$    &$\phantom{3}2.11 \pm0.03$  \\
15  &$3112.62\pm0.09$    &$\phantom{3}1.72 \pm0.02$  \\
16  &$3120.44\pm0.12$    &$\phantom{3}1.26 \pm0.02$  \\ \hline
\end{tabular}
\end{center}
\end{table}

A {\sc geant4}-based~\cite{geant4} simulation software package including
a description of the geometry and material and the detector response is used to generate Monte Carlo (MC) samples.
The {\sc babayaga}~\cite{babayaga} generator which includes interference between $A_{\rm cont}$ and $A_{\gamma}$ is used to simulate the $\elp\elm\to\mup\mum$ and the $\elp\elm\to\elp\elm$ events.
%{The generators for $\elp\elm\to 5\pi$ should describe all the intermediate states and their interferences.
%It requires sophisticated parametrization, which in principle, could be different on the top of $\jpsi$ and off
%at nearby energies, with parameters tuned by experimental data, which have not yet been available in sufficient statistics.
%Thus, we use a compromised way.}
{The samples $\elp\elm\to 5\pi$ and intermediate processes $\elp\elm\to\rhoz\rho^{\pm}\pi^{\mp}$, $\elp\elm\to\rhoz f_{2}(1275)\piz$, $\elp\elm\to\omega\pip\pim$ and $\elp\elm\to\eta\pip\pim$ are generated assuming a uniform phase space distribution.
The intermediate decays $\elp\elm\to\eta\rhoz$ and $\eta\omega$ and the subsequent decay of all intermediate states
are generated with {\sc evtgen}~\cite{evtgen2,evtgen3}.
For the $5\pi$ system, the polar angular distributions for each of the pions in the $\elp\elm$ CM frame are tuned to be the same as those in data.}
The {\sc mcgpj}~\cite{mcgpj} generator is used to incorporate radiation effects in the $\elp\elm\to 5\pi$ process.
The possible interference between $A_{g}$ and $A_{\rm cont}$ (or $A_{\gamma}$) is included in the {\sc mcgpj} generator.
{The output cross section from the {\sc mcgpj} generator is tuned to be the same as the observed cross section of $\elp\elm\to5\pi$.}
A MC sample of $\jpsi$ inclusive decays is used to explore possible hadronic background.
In this sample, the known decay modes are generated with {\sc evtgen} incorporating the branching fractions from the Particle Data Group
(PDG)~\cite{pdg} and the remaining unknown decays are generated according to the {\sc lundcharm}~\cite{lundcharm} model.
The CM energy spread is incorporated in all MC samples.

\section{ANALYSIS} \label{analysis}

\subsection{\bf Event selection for the \mbox{\boldmath $\elp\elm\to\mup\mum$} process}
Events of $\elp\elm\to\mup\mum$ are required to have only two charged tracks with opposite charge.
The charged tracks are required to originate from the interaction
region which is defined as a cylinder with a radius of $1$~cm and an axial
distance from the interaction point of $\pm10$~cm.
The polar angle $\theta$ of each track with respect to the positron beam
is required to be within the barrel region ($|\cos\theta|<0.8$).
Each charged track must have hit information in the EMC, and its measured energy deposit
divided by its momentum obtained from the MDC ($E/p$) is required
to be less than $0.3$ to suppress $\elp\elm\to\elp\elm$ and hadronic final state events.
\textcolor{black}{Cosmic rays} are rejected by requiring $\Delta T\equiv |T_{\rm trk1}-T_{\rm trk2}|<4$~ns,
where $T_{\rm trk1}$ and $T_{\rm trk2}$ are the measured flight times
in the TOF detector for the two tracks.
The improved track parameters obtained from the vertex fit,
which constrains the two tracks \textcolor{black}{to} a common vertex,
are used in further analysis. The momenta of muon candidates must satisfy
$(p_{\rm the}-4\sigma_{p})<p_{\mu^{\pm}}<(p_{\rm the}+3\sigma_{p})$, where
$p_{\rm the}$ and $\sigma_{p}$ are the nominal value and experimental resolution of the momentum of $\mu^{\pm}$, respectively.
Figure~\ref{fig_mass} (a) shows the momentum distributions of data and {the} {\sc babayaga} MC sample at $W=3092.32$~MeV.
Throughout this letter, all the performance plots are from the same energy point.
The data appear to be consistent with the MC simulations.

Potential two-body decay backgrounds are estimated by investigating the exclusive MC samples of
$\elp\elm\to\prp\prm$, $\kp\km$, $\pip\pim$, and $\elp\elm$. Only the process $\elp\elm\to\pip\pim$ is found to be a potential background.
According to Ref.~\cite{babar_2pi}, the cross section of $\pip\pim$ is about $10^{-2}$~nb
at $3000$~MeV, which is negligible compared {to that} of $\mup\mum$ of about $10$~nb.
At the $\jpsi$ peak, the ratio between the branching fraction of $\jpsi\to\pip\pim$ and that of $\jpsi\to\mup\mum$ is about $0.2\%$.
Taking into account the selection efficiency, where that of $\elp\elm\to\pip\pim$ is about one third of that of $\elp\elm\to\mup\mum$, the background from the $\pip\pim$ final state \textcolor{black}{can safely be} ignored. From a study of the $\jpsi$ inclusive MC sample,
the contribution from the remaining {multihadron} events is about $0.2$\% of the surviving events and is also negligible.

\subsection{\bf Event selection for the \boldmath{$\elp\elm\to5\pi$} and \boldmath{$\eta\pip\pim$} processes}
The events are required to have four charged tracks with a net charge of zero and at least two photons.
The charged tracks are required to originate from the interaction region,
while their polar angles are required to be within a range of $|\cos\theta|<0.93$.
Charged particle identification is performed by combining the ionization energy loss
($dE/dx$) in the MDC and the flight times in the TOF.
For each track, the probability for the pion particle hypothesis is required to be larger than that for the kaon particle hypothesis.
The photons are required to have a deposited energy
greater than 50 MeV in the endcap ($0.86<|\cos\theta|<0.92$) or 25 MeV in the barrel ($|\cos\theta|<0.8$) of the EMC.
To suppress electronic noise and energy deposits unrelated to the event, the time of the cluster signal given by the EMC must be within 700 ns after
the reconstructed event start time. To exclude clusters originating from charged tracks,
the angle between the photon candidate and the nearest charged track is required to be greater than $10^{\circ}$.

\begin{figure}[htbp]
\begin{center}
\includegraphics[angle=0,width=6.5cm, height=5.5cm]{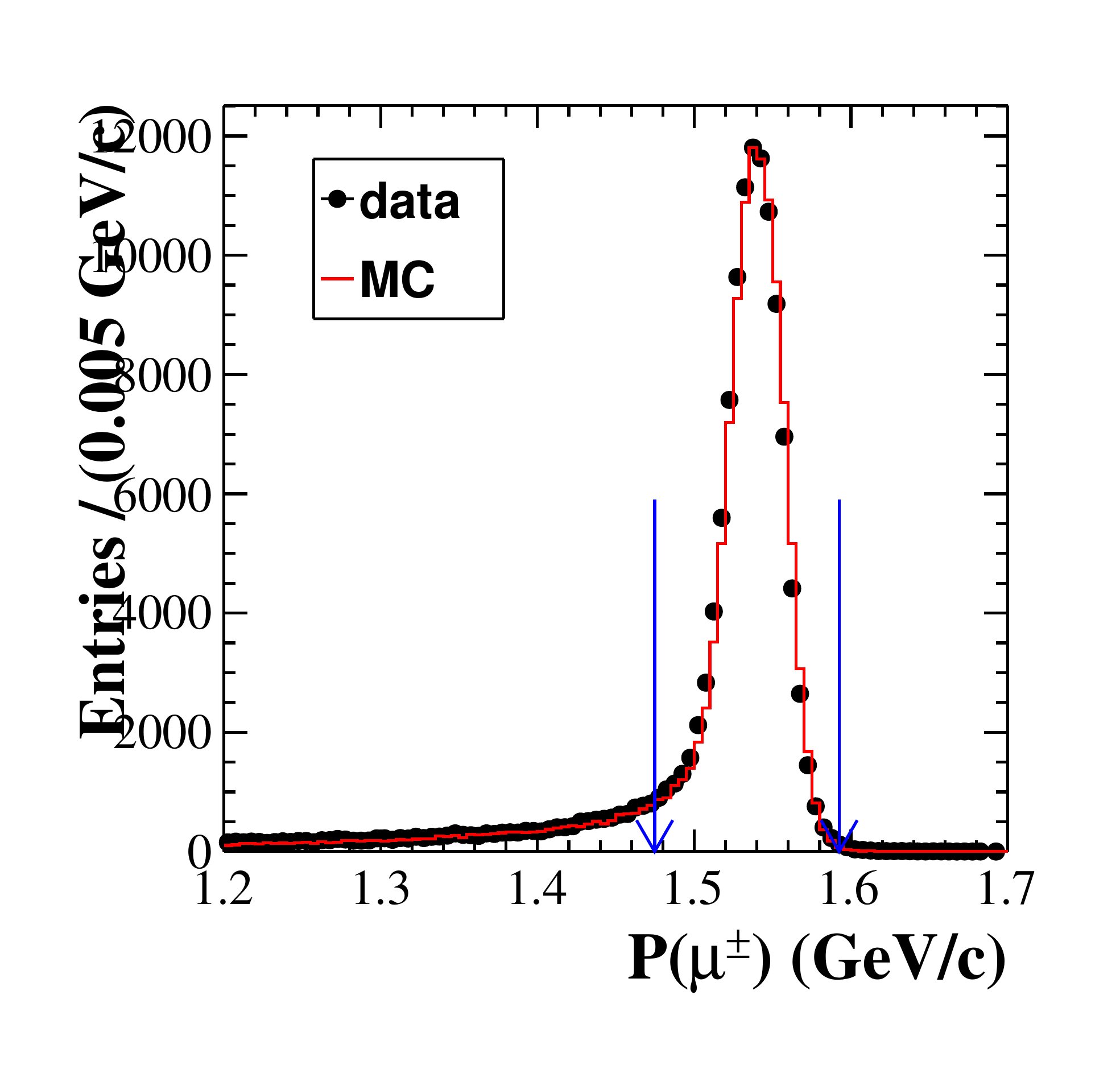}  \put(-120,80){(a)}
\includegraphics[angle=0,width=6.5cm, height=5.5cm]{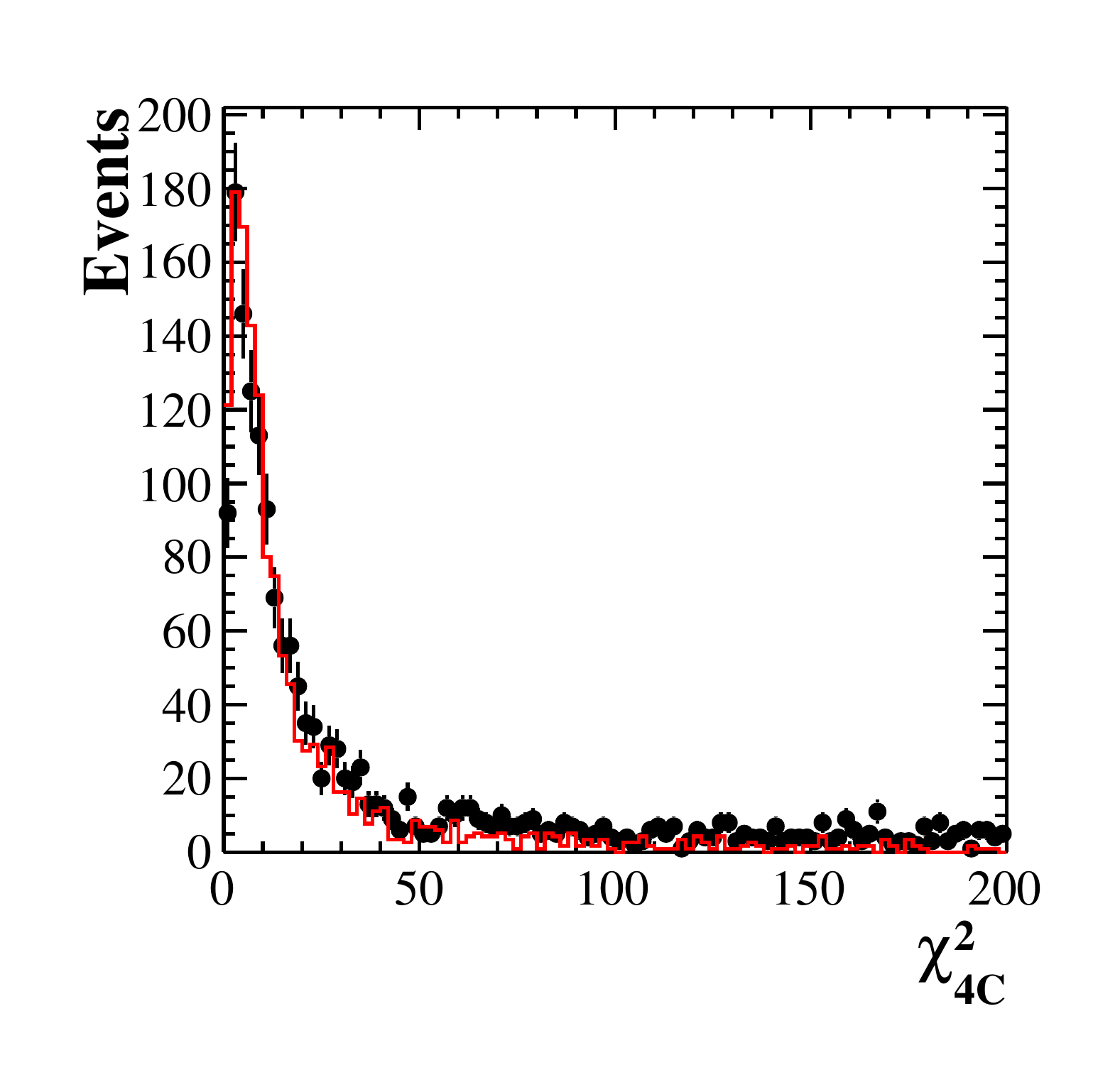}  \put(-120,80) {(b)} \\
\includegraphics[angle=0,width=6.5cm, height=5.5cm]{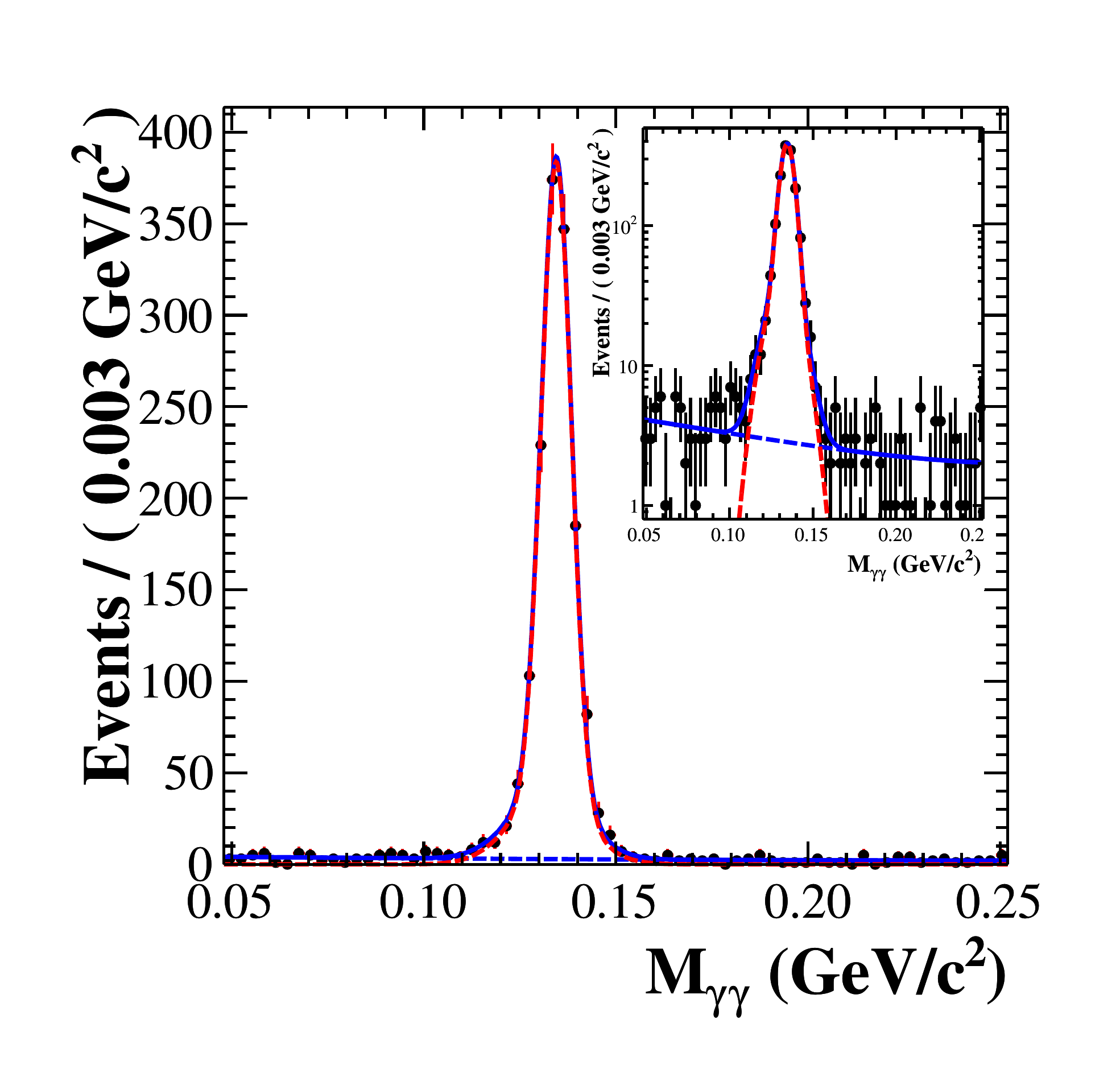}  \put(-120,80){(c)}
%\subfigure[]{\includegraphics[width=3.0cm,height=3cm]{fit_mpi0_2.pdf}}
%\begin{overpic}[scale=0.15]{fit_mpi0_2.pdf}\end{overpic}
\includegraphics[angle=0,width=6.5cm, height=5.5cm]{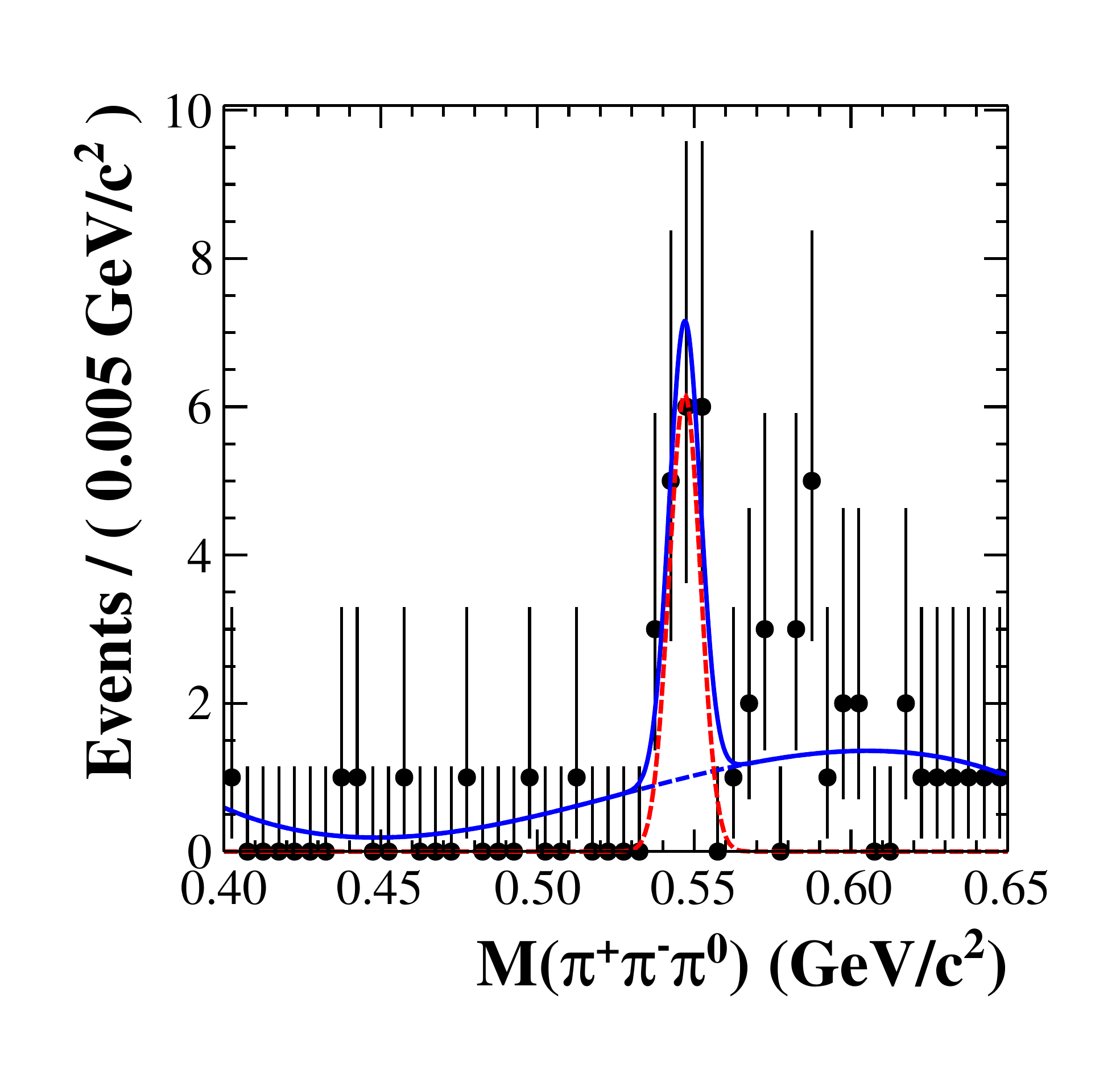}  \put(-120,80) {(d)}
\caption{\small (Colored online) (a) Comparison between data and {\sc babayaga} MC sample for the momentum of the $\mu^{\pm}$ for $\elp\elm\to\mup\mum$ candidate events. The region between the arrows denotes the signal window. (b) Comparison between data and MC simulation for the $\chi^{2}_{\rm 4C}$ in the $\elp\elm\to 5\pi$ channel. (c) The fit of the $M_{\gamma\gamma}$ spectrum, and the inset is in the logarithmic scale. (d) The fit of the $M^{\eta}_{\pip\pim\gamma\gamma}$ spectrum.
All plots are taken at $W=3092.32$~MeV, and the black dots with error bars are for data. For (a) and (b), the red histograms denote the MC samples. For (c) and (d), the red dashed lines denote the signal, the blue dotted lines are for the background, the blue solid lines represent the overall fit curve.}
\label{fig_mass}
\end{center}
\end{figure}

After constraining the four charged tracks to a common vertex using a vertex fit,
a four-constraint (4C) kinematic fit imposing energy and momentum conservation is performed
to the $\elp\elm\to2(\pip\pim)\gamma\gamma$ hypothesis.
Events with $\chisq{4\rm{C}}<200$ are retained, and at least 80\% of the background is rejected and about 95\% signal is retained.
If there are more than two photons, all combinations of photon pairs are tried and that with the least $\chisq{4\rm{C}}$ value is retained.
Figure~\ref{fig_mass} (b) shows the distribution of $\chisq{4\rm{C}}$ for the data and MC simulation.
The invariant mass of the photon pair $M_{\gamma\gamma}$ is required to be within the range $(0.0,0.3)$~GeV$/c^{2}$.
{
The decay angle ($\theta_{\rm decay}$) of a photon is defined as the polar angle measured in
the $\piz$ rest frame with respect to the $\piz$ direction
in the $\elp\elm$ CM frame.
The cosine of the decay angle ($\cos\theta_{\rm decay}$) is required to be lower than $0.9$}
%The decay angle of a photon in the $\piz$ rest frame $\theta_{\rm decay}$, is required to satisfy
%$\cos\theta_{\rm decay}\equiv\frac{|E_{\gamma_{1}}-E_{\gamma_{2}}|}{p_{\piz}}<0.9$ to
%The decay angle of a photon is the polar angle measured in the $\piz$ rest frame. % with respect to the $\piz$ direction
%in the $\elp\elm$ rest frame.  polar angle of a photon in the 0 rest frame,
%The cosinus of the decay angle $\cos\theta_{\rm decay}\equiv\frac{|E_{\gamma_{1}}-E_{\gamma_{2}}|}{p_{\piz}}$ is required to be less than $0.9$ to
to remove wrong photon combinations.%, where $E_{\gamma_{1,2}}$ is the energy deposited in the EMC of $\gamma_{1,2}$,
%and $p_{\piz}$ is the momentum of the $\piz$.
% which 4-momentum has been reconstructed from the momenta of $\gamma_{1}$ and $\gamma_{2}$~\cite{a0f0} in laboratory frame.

By studying the inclusive and exclusive MC samples, the backgrounds can be classified into
$\elp\elm\to\gamma2(\pip\pim)$, $\gamma\fivepi$ and $2(\pip\pim\piz)$ (abbreviated as $\gamma4\pi$, $\gamma5\pi$, and $6\pi$)
according to the number of photons in the final states.
{For normalization, the background channels are normalized according to their branching fractions from $\jpsi$~\cite{pdg}
decay or their {energy-dependent} cross section measured by BaBar~\cite{babar_6pi}.
Only the $\elp\elm\to\gamma 5\pi$ makes a peaking background of less than 1\% of the $\piz$ events on the spectrum of $M_{\gamma\gamma}$.}

The surviving candidate events include events from the process with an $\eta$ intermediate state,
i.e. $\elp\elm\to\eta\pip\pim$ with $\eta$ decays to $\pip\pim\piz$.
Due to G-parity conservation, {the dominant process $\elp\elm\to\eta\rhoz\to\eta\pip\pim$ is allowed only via EM decay}, and will
affect the measurement of $\Phi_{g,\rm EM}$ for the process $\elp\elm\to5\pi$. Thus, the process of $\elp\elm\to\eta\pip\pim$ will be separated
from the inclusive $\elp\elm\to 5\pi$, and measured alone.
In the inclusive $\elp\elm\to 5\pi$ candidate events, we reconstructed the $\eta$ signal with the $\pip\pim\gamma\gamma$ combination whose invariant
mass $M^{\eta}_{\pip\pim\gamma\gamma}$ is closest to the $\eta$ nominal mass.
The signal candidate of $\elp\elm\to 5\pi$ is then selected by imposing a further requirement of
$M^{\eta}_{\pip\pim\gamma\gamma}<0.517$~MeV$/c^{2}$ or $M^{\eta}_{\pip\pim\gamma\gamma}>0.577$~MeV$/c^{2}$.
The corresponding yield is determined by fitting the distribution of $\gamma\gamma$ invariant mass, $M_{\gamma\gamma}$, with
a double Gaussian function for the signal and a {second-order} polynomial function for background, as shown in Fig.~\ref{fig_mass} (c).
The yield of $\elp\elm\to\eta\pip\pim$ is determined by fitting the $M^{\eta}_{\pip\pim\gamma\gamma}$ distribution,
where the $\eta$ signal is modeled by a Gaussian function and the background is described by a {third- or lower-order} polynomial function,
as presented in Fig.~\ref{fig_mass} (d).
To better describe the data, the parameters of the $\eta$ and $\piz$ signal lineshapes are fixed to values obtained from fits to
distributions summed over all CM energies.

\subsection{\bf Cross sections of \boldmath{$\elp\elm\to\mup\mum$}, \boldmath{$5\pi$} and \boldmath{$\eta\pip\pim$}}

The observed cross section is calculated with
\begin{equation}
\sigma^{\rm obs}_{i} = \frac{N_{i}}{\epsilon_{i} \times \mathcal{L}_{i}(\times \mathcal{B})}, \notag
\end{equation}
where $N_{i}$ is the number of observed signal events, $\epsilon_{i}$ is the efficiency
given by the MC simulations, and $\mathcal{L}_{i}$ is the luminosity listed in Table~\ref{table_data set}. In the equation, $\mathcal{B}$
denotes the branching fractions of intermediate decays, and is
$\mathcal{B}(\piz\to\gamma\gamma)$ for $\elp\elm\to5\pi$ and
$\mathcal{B}(\eta\to\pip\pim\piz)\times\mathcal{B}(\piz\to\gamma\gamma)$
for $\elp\elm\to\eta\pip\pim$.
For $\elp\elm\to\mup\mum$, the efficiency from the {\sc babayaga} simulation includes the \textcolor{black}{radiative effects}~\cite{babayaga}.
%{For $\elp\elm\to 5\pi$, $\epsilon_{i}=f^{\rm EC}_{i}\times\epsilon^{\rm com}_{i}$ is used to take into
%account the radiation effect, where $f^{\rm EC}_{i}$ is an efficiency correction factor which is defined as the ratio of the detection
%efficiency of {\sc mcgpj} sample at the $i$-th energy point to the detection efficiency of {\sc mcgpj} sample at the $\jpsi$ energy.
%To take into account various kinematic effects of all the intermediate processes,
%the weighted averaged efficiency $\epsilon^{\rm com}_{i}$ is used,
%with weights corresponding to their relative production rates.
%Considering the similar detection efficiencies, the efficiency of
%$\elp\elm\to5\pi$ decaying in PHSP mode ($\epsilon^{\rm PHSP}_{i}$) is used for intermediate processes,
%such as $\elp\elm\to\rhoz\rho\pi$, $f_{2}(1270)\rho\pi$ and $a_{2}\pi\pi$, except
%for the process $\elp\elm\to\omega\pip\pim$ which has a lower efficiency.
%The combined efficiency is then approximated as
%$\epsilon^{\rm com}_{i}=\epsilon^{\omega\pip\pim}_{i}\times R^{\omega\pip\pim}_{i} + \epsilon^{\rm PHSP}_{i}\times (1-R^{\omega\pip\pim}_{i})$,
%where $R^{\omega\pip\pim}_{i}$ is the relative production ratio of $\mathcal{B}(\elp\elm\to\omega\pip\pim)$ to
%$\mathcal{B}(\elp\elm\to 5\pi)$, calculated as
%$(\frac{N^{\omega\pip\pim}_{i}}{\epsilon^{\omega\pip\pim}_{i}})/(\frac{N^{5\pi}_{i}}{\epsilon^{\rm PHSP}_{i}})$.}

{For the process $\elp\elm\to 5\pi$, to take into account kinematic effects of the intermediate states,
the weighted-average efficiency $\epsilon^{\rm com}_{i}$ obtained according to the relative production rates between
the processes with different intermediate states is used.}
{The interference among different intermediate processes is assumed to be independent of the phase measurement and not taken into account.}
To take into account the radiation effect, an additional CM {energy-dependent} correction factor,
$f^{\rm EC}_{i}$ is used, which is the ratio of the detection efficiencies of $\elp\elm\to 5\pi$ at the
$i$-th CM energy point estimated with the generator {\sc mcgpj} to that at the $\jpsi$ peak.
{The generator {\sc mcgpj} models radiation effect for the process $\elp\elm\to 5\pi$ properly
by adjusting the output cross section to be the same as the calculated $\sigma^{\rm obs}_{i}$ from data}.
Thus, the effective detection efficiency is $\epsilon_{i}=f^{\rm EC}_{i}\times\epsilon^{\rm com}_{i}$.

{From the PDG, we know the decays $\jpsi\to\eta\rho$, $\eta\omega$, and $\eta\pip\pim$ also exist,
even though the measured branching ratios are very old and have large uncertainties.
According to MC simulations, the efficiencies for these processes are nearly the same.
Thus, the efficiency of the MC sample for $\elp\elm\to\eta\pip\pim$, without intermediate states, is used in the cross section calculation.}
The efficiency correction factor $f^{\rm EC}_{i}$ is not implemented due to the large statistical uncertainty {of its cross section} and the small effect of $f^{\rm EC}_{i}$ on the phase measurement (see the results of the $5\pi$ in Section~\ref{results}).
The calculated cross sections for $\elp\elm\to\mup\mum$, $5\pi$ and $\eta\pip\pim$,
together with the efficiencies and the number of events, are listed in Table~\ref{table_evt}.

\begin{table}[th!]
%\begin{center}
\caption{\small The number of events, efficiency and the observed cross section
for $\elp\elm\to\mup\mum$, $5\pi$ and $\eta\pip\pim$ at each energy point.
Statistical uncertainties are quoted for the number of events and the efficiencies,
while both statistical and systematic uncertainties are quoted for the cross section.
}
\label{table_evt}
\begin{scriptsize}
\centering
\begin{tabular}{c|r@{$\pm$}lcr@{$\pm$}c@{$\pm$}l|r@{$\pm$}lcr@{$\pm$}c@{$\pm$}l} \hline
&\multicolumn{6}{c|}{$\mup\mum$}     &\multicolumn{6}{c}{$5\pi$} \\
No. &\multicolumn{2}{c}{$N_{i}$}  &$\epsilon_{i}$(\%)   &\multicolumn{3}{c|}{$\sigma^{\rm obs}_{i}$~(nb)}   &\multicolumn{2}{c}{$N_{i}$}    &$\epsilon_{i}(\%)$  &\multicolumn{3}{c}{$\sigma^{\rm obs}_{i}$~(nb)}           \\\hline
1   &$ 76553$ &$277$   &$54.52\pm0.16$   &$9.411 $ &$0.034$ &$0.217$     &$734  $ &$29 $   &$23.60\pm 1.28$   &$0.211$ &$0.008$ &$0.017$ \\
2   &$ 76058$ &$276$   &$54.53\pm0.16$   &$9.261 $ &$0.034$ &$0.213$     &$723  $ &$28 $   &$23.88\pm 1.43$   &$0.204$ &$0.008$ &$0.017$ \\
3   &$ 81532$ &$286$   &$53.30\pm0.16$   &$8.794 $ &$0.031$ &$0.202$     &$765  $ &$29 $   &$23.54\pm 1.25$   &$0.189$ &$0.007$ &$0.015$ \\
4   &$ 21584$ &$147$   &$53.74\pm0.16$   &$8.42  $ &$0.06 $ &$0.20 $     &$180  $ &$14 $   &$24.31\pm 3.02$   &$0.158$ &$0.012$ &$0.021$ \\
5   &$ 63674$ &$252$   &$52.76\pm0.16$   &$7.758 $ &$0.031$ &$0.177$     &$858  $ &$30 $   &$25.16\pm 1.27$   &$0.222$ &$0.008$ &$0.017$ \\
6   &$ 51677$ &$227$   &$51.12\pm0.16$   &$6.780 $ &$0.030$ &$0.155$     &$1434 $ &$39 $   &$26.09\pm 1.02$   &$0.373$ &$0.010$ &$0.027$ \\
7   &$ 15929$ &$126$   &$58.84\pm0.16$   &$12.63 $ &$0.10 $ &$0.30 $     &$4962 $ &$71 $   &$28.69\pm 0.60$   &$ 8.16$ &$0.12 $ &$0.53 $ \\
8   &$ 52001$ &$228$   &$63.23\pm0.17$   &$45.28 $ &$0.20 $ &$1.07 $     &$18120$ &$140$   &$28.37\pm 0.40$   &$35.59$ &$0.27 $ &$2.26 $ \\
9   &$154741$ &$393$   &$63.87\pm0.15$   &$113.47$ &$0.29 $ &$2.67 $     &$52380$ &$230$   &$28.42\pm 0.35$   &$ 87.4$ &$0.4  $ &$5.5  $ \\
10  &$281713$ &$531$   &$63.99\pm0.16$   &$212.8 $ &$0.4  $ &$ 5.1 $     &$90560$ &$310$   &$28.19\pm 0.31$   &$157.1$ &$0.5  $ &$9.9  $ \\
11  &$155118$ &$394$   &$64.07\pm0.16$   &$109.90$ &$0.28 $ &$2.60 $     &$43520$ &$210$   &$28.32\pm 0.36$   &$70.57$ &$0.34 $ &$4.47 $ \\
12  &$ 26646$ &$163$   &$62.62\pm0.15$   &$56.29$  &$0.35 $ &$1.39 $     &$6424 $ &$81 $   &$28.41\pm 0.52$   &$ 30.3$ &$0.4  $ &$2.0  $ \\
13  &$ 21893$ &$148$   &$60.51\pm0.15$   &$22.44$  &$0.15 $ &$0.54 $     &$3440 $ &$60 $   &$26.57\pm 0.68$   &$ 8.13$ &$0.14 $ &$0.54 $ \\
14  &$ 20184$ &$142$   &$58.74\pm0.16$   &$16.32$  &$0.12 $ &$0.38 $     &$2468 $ &$50 $   &$27.89\pm 0.79$   &$ 4.25$ &$0.09 $ &$0.29 $ \\
15  &$ 13173$ &$115$   &$57.72\pm0.16$   &$13.27$  &$0.12 $ &$0.32 $     &$1160 $ &$35 $   &$26.72\pm 1.11$   &$ 2.55$ &$0.08 $ &$0.19 $ \\
16  &$ 8550 $ &$93$    &$56.40\pm0.16$   &$11.99$  &$0.13 $ &$0.29 $     &$623  $ &$26 $   &$26.63\pm 1.43$   &$ 1.87$ &$0.08 $ &$0.15 $ \\\hline
\end{tabular}
\begin{tabular}{c|r@{$\pm$}lcr@{$\pm$}c@{$\pm$}l}
&\multicolumn{6}{c}{$\eta\pip\pim$}   \\
No.   &\multicolumn{2}{c}{$N_{i}$}   &$\epsilon_{i}$(\%)   &\multicolumn{3}{c}{$\sigma^{\rm obs}_{i}$~(nb)}   \\\hline
1   &$32 $ &$ 6$  &$21.16\pm 0.11$   &$0.045 $ &$0.009 $ &$0.006$ \\
2   &$24 $ &$ 6$  &$21.08\pm 0.11$   &$0.034 $ &$0.008 $ &$0.004$ \\
3   &$34 $ &$ 6$  &$20.78\pm 0.10$   &$0.042 $ &$0.008 $ &$0.006$ \\
4   &$8  $ &$ 3$  &$21.07\pm 0.11$   &$0.037 $ &$0.015 $ &$0.005$ \\
5   &$25 $ &$ 6$  &$21.11\pm 0.11$   &$0.033 $ &$0.007 $ &$0.004$ \\
6   &$15 $ &$ 5$  &$21.14\pm 0.11$   &$0.0216$ &$0.0064$ &$0.0025$ \\
7   &$10 $ &$ 4$  &$21.25\pm 0.11$   &$0.100 $ &$0.039 $ &$0.013$ \\
8   &$19 $ &$ 7$  &$20.94\pm 0.11$   &$0.218 $ &$0.076 $ &$0.027$ \\
9   &$60 $ &$11$  &$21.00\pm 0.11$   &$0.59  $ &$0.11  $ &$0.07$ \\
10  &$118$ &$15$  &$20.79\pm 0.10$   &$1.21  $ &$0.15  $ &$0.15$ \\
11  &$74 $ &$11$  &$20.83\pm 0.10$   &$0.709 $ &$0.105 $ &$0.088$ \\
12  &$22 $ &$ 6$  &$20.50\pm 0.10$   &$0.63  $ &$0.16  $ &$0.08$ \\
13  &$12 $ &$ 4$  &$20.84\pm 0.10$   &$0.155 $ &$0.056 $ &$0.020$ \\
14  &$7  $ &$ 3$  &$20.71\pm 0.10$   &$0.072 $ &$0.034 $ &$0.009$ \\
15  &$5  $ &$ 3$  &$20.58\pm 0.10$   &$0.057 $ &$0.036 $ &$0.007$ \\
16  &$6  $ &$ 3$  &$20.63\pm 0.10$   &$0.094 $ &$0.045 $ &$0.012$ \\\hline
%\\\hline
\end{tabular}
\end{scriptsize}
%\end{center}
\end{table}

\subsection{\bf Systematic uncertainties}

Systematic uncertainties are divided into two categories.  Those that are universal among the
different energy points include
those related to the event selection efficiencies, {intermediate states in $\elp\elm\to\eta\pip\pim$},
and the branching fractions of intermediate state decays.
Those that are not universal are treated separately for all energy points, which include
the uncertainties related to the fits to the spectra, %$f^{\rm EC}_{i}$,
{$\epsilon^{\rm com}_{i}$ of $\elp\elm\to 5\pi$}, and the luminosities.

The systematic uncertainty of the tracking of muons is studied with a control sample of $\jpsi\to\mup\mum$
selected with more stringent criteria on one tagged charged track.
The efficiency is the rate to detect another charged track on the recoil side of the tagged track.
The difference on the efficiency is 1\% between data and MC simulation, which is treated as the systematic uncertainty.
The systematic uncertainties associated with the tracking and the particle identification for pion candidates are investigated
using a control sample of $\jpsi\to\prp\prm\pip\pim$, and are found to be 1\% individually~\cite{yuanwl}.
Dedicated studies on $\elp\elm\to\gamma\mup\mum$~\cite{vindy} and $\jpsi\to\pip\pim\piz$~\cite{a0f0}
conclude that the systematic uncertainty due to photon identification is $1\%$ per photon.
The systematic uncertainty related to the 4C kinematic fit is determined by changing the $\chisq{\rm 4C}$ requirement, and found to be $1\%$.
{The uncertainties of the branching fractions for the intermediate-state decays} $\piz\to\gamma\gamma$ and
$\eta\to\pip\pim\piz$ from the PDG~\cite{pdg} are considered in the systematic uncertainty.

The requirements of $\cos\theta$, $E/p$, $|\Delta T|$ and $p_{\mu^{\pm}}$ in the selection of $\elp\elm\to\mup\mum$,
and $M^{\eta}_{\pip\pim\gamma\gamma}$ and $\cos\theta_{\rm decay}$ in the selection of $\elp\elm\to5\pi$
%and $\cos\theta_{\rm decay}$ in the selection of $\elp\elm\to\eta\pip\pim$
are varied at all energy points.
The largest difference of the cross section with respect to the nominal result at each energy point is taken as the deviation of each requirement.
The {weighted-average deviation} (with weights of statistics of each energy point) of each item is taken as the uncertainties.
The uncertainties of $\cos\theta$, $E/p$, $|\Delta T|$, $p_{\mu^{\pm}}$, $M^{\eta}_{\pip\pim\gamma\gamma}$
and $\cos\theta_{\rm decay}$ are determined as 0.16\%, 0.09\%, 0.05\%, 0.26\%, 0.04\%, and 0.40\%, respectively.
The uncertainties of the requirement of $\cos\theta_{\rm decay}$ are the same for the processes of
$\elp\elm\to 5\pi$ and $\elp\elm\to\eta\pip\pim$.

The uncertainties associated with the fit procedure on the $M_{\gamma\gamma}$ and
$M^{\eta}_{\pip\pim\gamma\gamma}$ distributions are estimated by changing
the signal shapes to the {Crystal Ball} function and MC simulated histograms, respectively,
extending or shrinking the fit ranges, changing the background shapes to a
higher or lower order of the polynomial functions, and changing the interval width of each spectrum.
The largest deviations of results for the different fit scenarios with respect to
the nominal values are regarded as the individual systematic uncertainties
and are added in quadrature to be the systematic uncertainty associated with the fit procedure.
Due to the low statistics in the process of $\elp\elm\to\eta\pip\pim$, ensembles of simulated
data samples (toy MC samples) at each energy point are generated according to the nominal
fit result with the same statistics as data, then fitted by the alternative fitting scenario.
These trials are performed 1000 times, and the average signal yields are taken as the results.
For the data with the CM energy being $3101.92$, $3106.14$, $3112.62$, and $3120.44$ MeV,
the statistics are extremely low and the {uncertainties} of the fit procedure are assigned
to be the same as that for data at CM energy {of} $3099.61$~MeV.
Totally, the fit procedure introduces systematic uncertainties of about $1-2\%$ and $11\%$
for the channels $\elp\elm\to5\pi$ and $\eta\pip\pim$, respectively.

{%From PDG, $\mathcal{B}(\jpsi\to\eta\pip\pim)=(4.0\pm1.7)\times10^{-4}$, $\mathcal{B}(\jpsi\to\eta\rhoz)=(1.93\pm0.23)\times10^{-4}$.
%and $\mathcal{B}(\jpsi\to a_{2}(1320)\pi)<4.3\times10^{-3}$.
The systematic uncertainty due to the intermediate states in $\elp\elm\to\eta\pip\pim$ is about $3.0\%$, estimated as the difference between
the weighted-average efficiency which takes into account the efficiencies and the relative branching fractions of $\jpsi\to\eta\pip\pim$ and $\jpsi\to\eta\rhoz$ and the efficiency of $\jpsi\to\eta\pip\pim$.}

%{The uncertainty of $f^{\rm EC}_{i}$ mainly comes from the statistical uncertainty of simulation.}
The uncertainty associated with $\epsilon^{\rm com}_{i}$ in the decay $\elp\elm\to 5\pi$ mainly comes from
the statistical uncertainty of {the relative ratios among different processes.} %$R^{\omega\pip\pim}_{i}$.
{Besides, the measured angular distributions of the pions in the MC samples are corrected to be the same as those measured in data.
The systematic uncertainty due to the correction is estimated to be 0.1\%-5.8\% depending on the statistics of each dataset.}
The uncertainty of the luminosity determination is determined to be 1.1-1.3\%, as listed in Table~\ref{table_data set}.

All the systematic uncertainties discussed above are
combined in quadrature to obtain the overall systematic uncertainties.

\section{RESULTS} \label{results}

Due to the effects of radiation and CM energy spread, the observed cross section cannot be directly compared with the Born cross section.
In this section, the fit formulas for the observed cross section and fit results are presented.

The Born cross section of $\elp\elm\to\mup\mum$, consisting of only the pure EM contributions,
$A_{\gamma}$ and $A_{\rm cont}$, is conventionally expressed as~\cite{ref_slc,ref_bes,ref_kder,BW}:
\begin{eqnarray}
\sigma^{0}(W)=\frac{4\pi\alpha^{2}}{W^{2}}\left|1+\frac{3W^{2}\sqrt{\Gamma_{ee}\Gamma_{\mu\mu}}e^{i\Phi_{\gamma,\rm cont}}}{\alpha M(W^{2}-M^{2}+iM\Gamma)}\right|^{2},\notag
\label{eq_mumu_BW}
\end{eqnarray}
where $\alpha$ is the fine structure constant, $M$ and $\Gamma$ are the mass and width of the $\jpsi$,
and $\Gamma_{ee}$ and $\Gamma_{\mu\mu}$ are the partial widths of $\jpsi\to\elp\elm$ and $\mup\mum$, respectively.
Incorporating the {radiative} correction $F(x,W)$, the cross section reads:
\begin{equation}
\label{eq_ISR2}
\sigma^{\prime}(W)=\int \limits_{0}^{1-(\frac{W_{\rm min}}{W})^2} dx F(x,W)\sigma^{0}(W\sqrt{1-x}), \notag
\end{equation}
where $W_{\rm min}$ is the minimum invariant mass of the $\mup\mum$ system,
$x=\frac{2E_{\gamma}}{\sqrt{s}}$, $E_{\gamma}$ is the energy of the radiation photon,
and $F(x,W)$ is approximated as ~\cite{Kuraev85}:
\begin{eqnarray}
\begin{split}
\label{eq_ISRKF}
&F(x,W)= x^{\beta-1}\beta \cdot (1+\delta)-\beta(1-\frac{x}{2})
+\frac{1}{8}\beta^{2}\left[4(2-x)\ln\frac{1}{x}-\frac{1+3(1-x)^2}{x}\ln(1-x)-6+x\right], \notag
\end{split}
\end{eqnarray}
with %$x=\frac{E_{\gamma}}{W^{2}}$ ($E_{\gamma}$ is the radiated energy),
$\delta=\frac{3}{4}\beta+\frac{\alpha}{\pi}(\frac{\pi^{2}}{3}-\frac{1}{2})+\beta^{2}(\frac{9}{32}-\frac{\pi^{2}}{12})$ and
$\beta=\frac{2\alpha}{\pi} (2\ln \frac{W}{m_{e}}-1)$.
The CM energy spread ($S_{E}$) is included by convolving with a Gaussian function with a width of $S_{E}$,
and the expected cross section $\sigma^{\prime\prime}(W_{i})$ at $W_{i}$ is obtained as
\begin{equation}
\label{eq_tot2}
\sigma^{\prime\prime}(W)
=\int^{W+nS_{E}}_{W-nS_{E}} \frac{1}{\sqrt{2\pi}S_{E}}\exp\left(\frac{-(W-W^{\prime})^{2}}{2S^{2}_{E}}\right)\sigma^{\prime}(W^{\prime})dW^{\prime}. \notag
\end{equation}
Finally, the minimizing function is built with the factored minimization method separating
the correlated and uncorrelated systematic uncertainties, and the effective {variance-weighted}
least squares method~\cite{effvar} including the uncertainty $\Delta W$ along the $X$-axis
by projecting it along the $Y$-axis. The resulting $\chi^{2}$ reads
\begin{eqnarray}
\label{eq_chisq}
\chi^{2} = \sum^{16}_{i=1} \frac{ \left[\sigma^{\rm obs}_{i}-f\sigma^{\prime\prime}(W_{i})\right]^{2} }{ (\Delta\sigma^{\rm obs}_{i})^{2} + \left[\Delta W_{i}\cdot \frac{d\sigma^{\prime\prime}(W)}{dW}\right]^{2} } + \left(\frac{1-f}{\Delta f}\right)^2, \notag
\end{eqnarray}
\textcolor{black}{where $f$} is the normalization factor, and $\Delta f$ is its uncertainty and set as the total correlated systematic {uncertainty}.
The term $\Delta\sigma^{\rm obs}_{i}$ is the combined statistical and uncorrelated systematic uncertainties.
Thus, the obtained uncertainties of results in this section include statistical and systematic ones.
To be more efficient in the fitting procedure, instead of two integrations, an approximation~\cite{wangp,fengzhi,zhouxy}
is used for $\sigma^{\prime}(W)$, which is decomposed into a resonance and interference term
$\sigma^{\rm R+I}(W)$ as well as a continuum term $\sigma^{\rm C}(W)$.
\textcolor{black}{The formulas} can be found in the Appendix.
The vacuum polarization is included by quoting the value of $\Gamma_{ee}$, $\Gamma_{\mu\mu}$ and
$\Gamma$ from the world averaged values~\cite{pdg,Alexander}.

\begin{figure}[htbp]
\begin{center}
\includegraphics[angle=0,width=5.5cm, height=5cm]{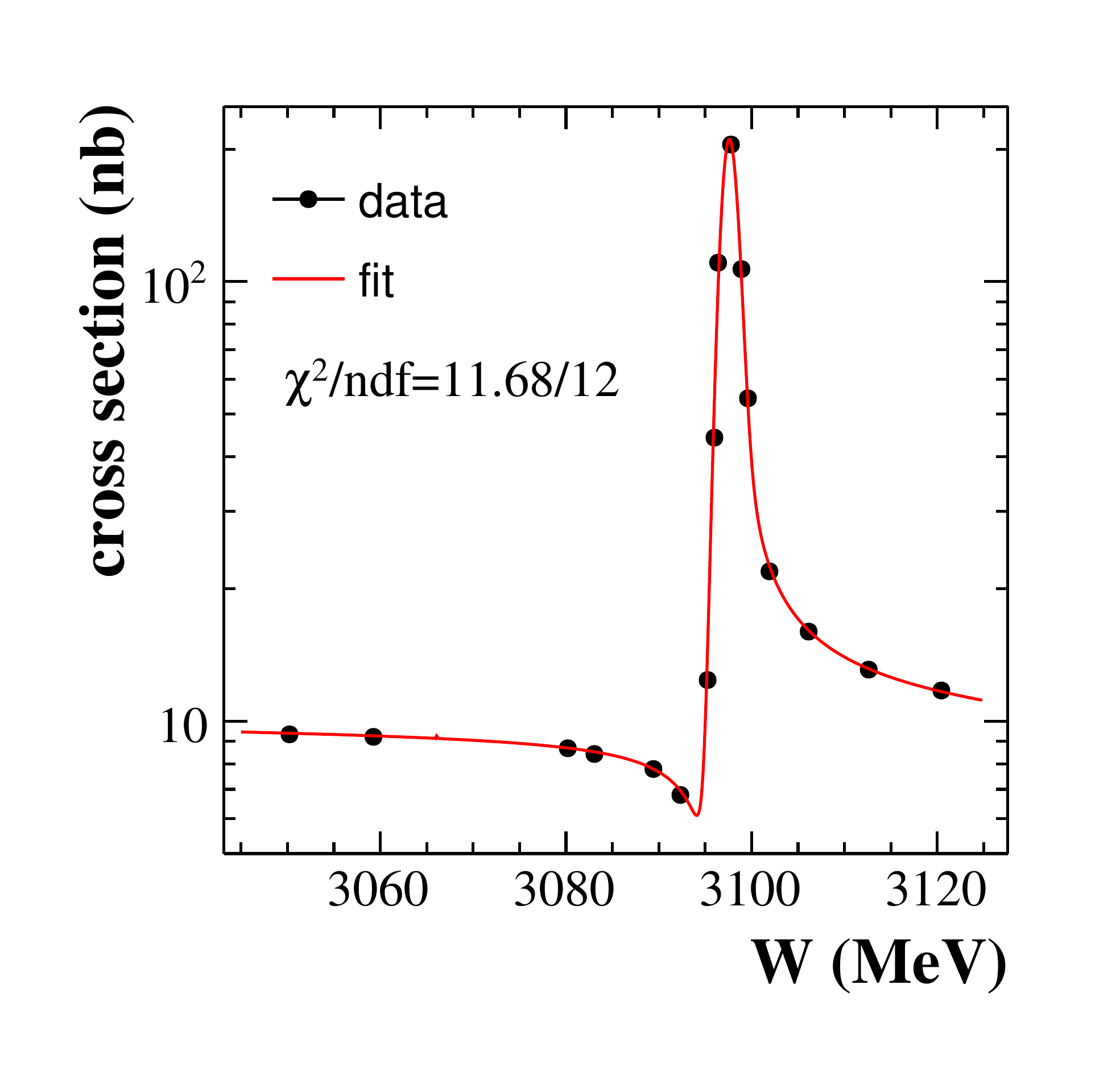}  \put(-100,65){(a)}
\includegraphics[angle=0,width=5.5cm, height=5cm]{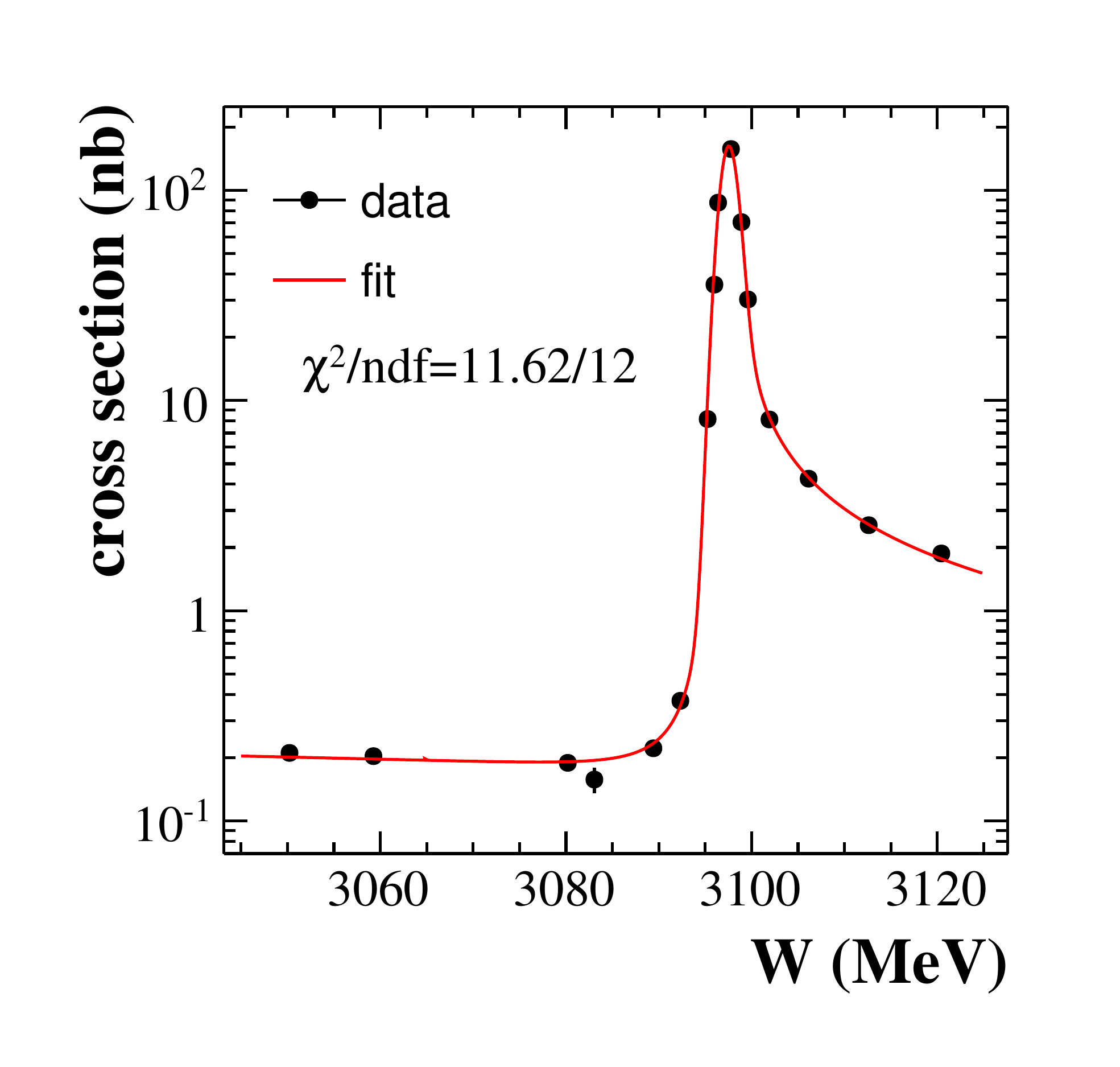}  \put(-100,65) {(b)}
\includegraphics[angle=0,width=5.5cm, height=5cm]{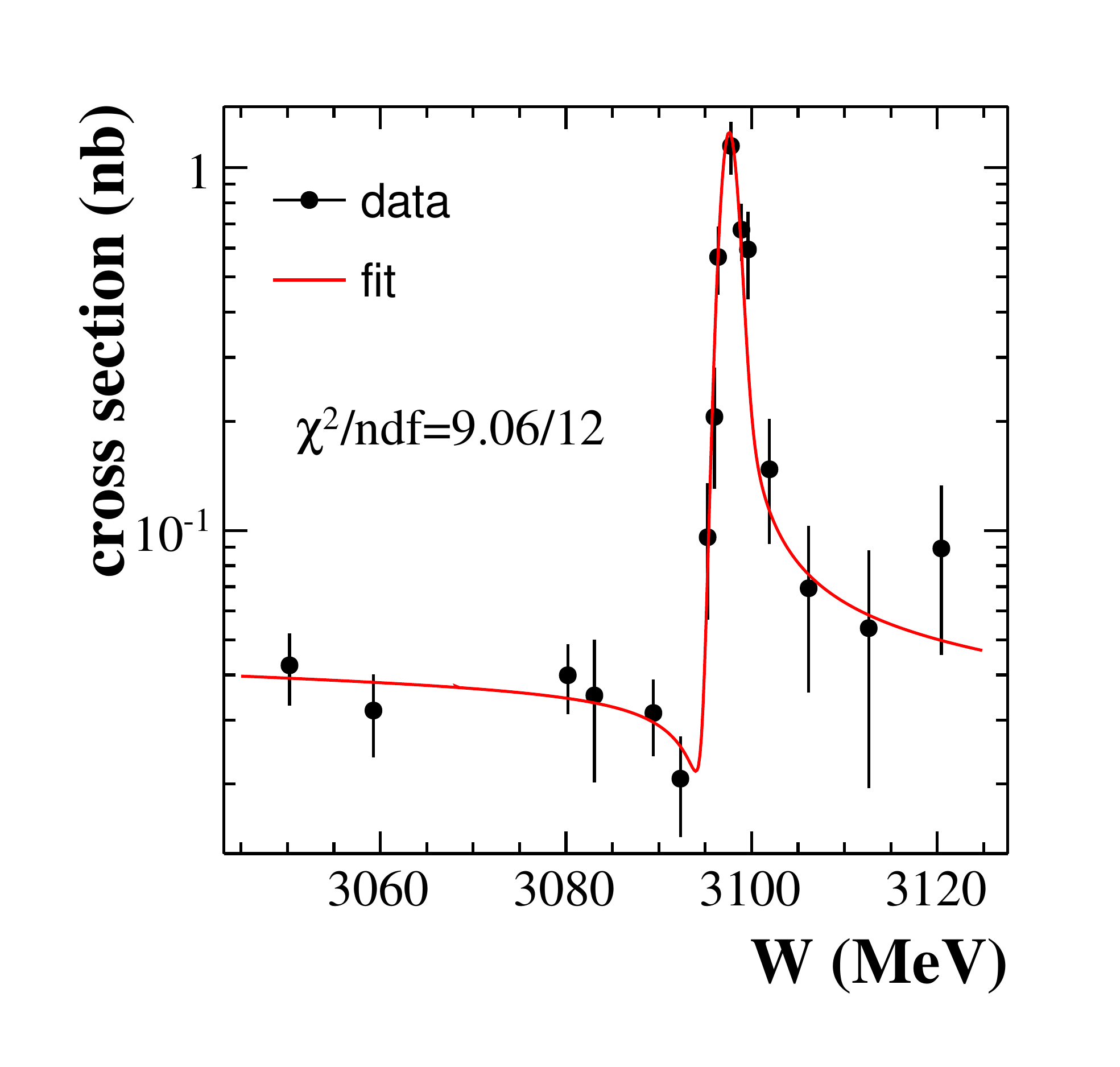} \put(-100,65) {(c)}
\caption{\small The lineshapes of $\elp\elm$ annihilates to (a) $\mup\mum$, (b) $5\pi$,
and (c) $\eta\pip\pim$. The black points with error bars are data,
and the solid lines show the fit results.}
\label{fig_5pi}
\end{center}
\end{figure}

We perform the minimized $\chi^{2}$ fit to the measured cross section of $\elp\elm\to\mup\mum$
with the free parameters $S_{E}$, $M$, and $\Phi_{\gamma,\rm cont}$
and the fit curve is presented in Fig.~\ref{fig_5pi} (a).
The minimization gives $\Phi_{\gamma,\rm cont}=(3.0\pm5.7)^{\circ}$, which is consistent with zero \textcolor{black}{as expected.}
A scan of the \textcolor{black}{parameter $\Phi_{\gamma,\rm cont}$} in the full range $(-180^{\circ},180^{\circ})$ confirms that only one solution exists.
By further fixing the parameter $\Phi_{\gamma,\rm cont}$ to be zero, an alternative fit is carried out,
\textcolor{black}{resulting in values of $S_{E}$ and $M$ consistent with the previous fit.}
The resultant $M$ is higher than its world average value
$M_{\jpsi}$~\cite{pdg}, and indicates a deviation of the absolute energy calibration for the BEMS.
Therefore, to \textcolor{black}{obtain} the proper detection efficiencies, the CM energies of MC samples are corrected by shifting
the absolute values with $\Delta M= M-M_{\jpsi}$.
To estimate the effect of fixed $\Gamma_{ee}$, $\Gamma_{\mu\mu}$ and $\Gamma$ in the fit,
the alternative fits are performed by letting these variables \textcolor{black}{free in the fit}.
We also perform two additional fits, \textcolor{black}{one with} an alternative analytical formula in
Ref.~\cite{ref_kder} which takes \textcolor{black}{a} different approximation, and the other is without \textcolor{black}{the} factored minimization method,
in which the $\Delta\sigma^{\rm obs}_{i}$ is the statistical and systematic uncertainties added in quadrature.
All of the results from various fits turn out to be consistent with each other.
Taking into account the differences among different approaches and parameterizations,
we \textcolor{black}{obtain} $\Phi_{\gamma,\rm cont}=(3.0\pm10.0)^{\circ}$, $S_{E}=(0.90\pm0.03)$~MeV
and $\Delta M=(0.57\pm0.05)$~MeV$/c^{2}$, which will be used in the fit for the hadronic final state.

By assuming the $\Phi_{\gamma,\rm cont}$ to be zero, the Born cross section for $\elp\elm\to5\pi$ is written as
\begin{eqnarray}
\label{eq_5pi}
\sigma^{0}(W)=\left(\frac{\mathcal{A}}{W^2}\right)^2\frac{4\pi\alpha^{2}}{W^2}
\left|1+\frac{3W^2\sqrt{\Gamma_{ee}\Gamma_{\mu\mu}}(1+\mathcal{C}e^{i\Phi_{g,\rm EM}})}{\alpha M(W^2-M^{2}+iM\Gamma)}\right|^{2}, \notag
\end{eqnarray}
where $\mathcal{C}$ is the ratio of $\frac{|A_{g}|}{|A_{\gamma}|}$, $\frac{\mathcal{A}}{W^2}$ is the form factor,
and $\mathcal{A}$ is a free parameter in the fits.
The decay width of $\jpsi\to 5\pi$ can be calculated as
$\Gamma_{5\pi}=(\frac{\mathcal{A}}{W^2})^2\Gamma_{\mu\mu}|\mathcal{C}e^{i\Phi_{g,\rm EM}}+1|^2$, and the corresponding branching fraction is then
$\mathcal{B}(\jpsi\to 5\pi)=\Gamma_{5\pi}/\Gamma$.
The analytical form $\sigma^{\prime}(W)$ for the $\elp\elm\to 5\pi$ is similar to that for $\elp\elm\to\mup\mum$ (see Appendix).
Similar to the fit for $\elp\elm\to\mup\mum$, we fix \textcolor{black}{the} parameters $\Gamma_{ee}$, $\Gamma_{\mu\mu}$ and $\Gamma$
\textcolor{black}{to the world average values}~\cite{pdg} in the fit.
By further fixing the values of $M$ and $S_{E}$ obtained in $\elp\elm\to\mup\mum$ and
constraining $\Phi_{g,\rm EM}$ to be within $(0,180)^{\circ}$,  the minimization fit yields
{$\Phi_{g,\rm EM}=(84.9\pm2.6)^{\circ}$} and {$\mathcal{B}(\jpsi\to 5\pi)=(4.73\pm0.41)\%$}.
An alternative fit with $M$ and $S_{E}$ \textcolor{black}{being free parameters} is performed to estimate the systematic uncertainty
associated with the fixed $M$ and $S_{E}$.
The {uncertainties} associated with those of $\Gamma_{ee}$, $\Gamma_{\mu\mu}$ and $\Gamma$ are evaluated
by shifting the corresponding values within one standard deviation.
The impact of $f^{\rm EC}_{i}$ is estimated by fitting the cross sections without incorporating it.
The minimization function without \textcolor{black}{the factor $f$} is also applied to consider its influence.
All results of $\Phi_{g,\rm EM}$ and $\mathcal{B}(\jpsi\to5\pi)$ are consistent within the uncertainty
and the differences {are included into the final uncertainties which listed in Table~\ref{table_5pi_fit}}.

\begin{table}[htpb]
\begin{center}
\caption{\small Fit results for the lineshape of $\elp\elm\to5\pi$.
Solution I is with $\Phi_{g,\rm EM}>0^{\circ}$ and solution II is with $\Phi_{g,\rm EM}<0^{\circ}$.}
\label{table_5pi_fit}
\begin{tabular}{cccccccccccc} \hline
            &$\Phi_{g,\rm EM}$                  &$\mathcal{B}_{5\pi}$\,(\%)  &{$\chi^{2}/\rm{ndf}$}  \\\hline
Solution I  &{$\phantom{1}(84.9\pm3.6)^{\circ}$}  &{$4.73\pm0.44$}  &{$11.62/12$}  \\
Solution II &{$(-84.7\pm3.1)^{\circ}$} &{$4.85\pm0.45$}  &{$11.62/12$} \\\hline
\end{tabular}
\end{center}
\end{table}

Constraining \textcolor{black}{$\Phi_{g,\rm EM}$} in the interval $(-180,0)^{\circ}$, the second solution is found.
The similar alternative fits described above are also carried out to estimate the corresponding systematic uncertainties.
The results are listed in Table~\ref{table_5pi_fit}, too.
%which are consistent within the uncertainties with respect to the first solution.
%{}
{As a check, the branching ratio of $\jpsi\to5\pi$ {\it via} only $A_{\gamma}$
can be calculated by $\mathcal{B}(\jpsi\to5\pi)\frac{1}{|1+\mathcal{C}|^2}$, and it is consistent with
the result calculated from $\mathcal{B}(\jpsi\to\mup\mum)\frac{\sigma(\elp\elm\to5\pi)}{\sigma(\elp\elm\to\mup\mum)}$ within uncertainty.}

The process $\elp\elm\to\eta\pip\pim$ is dominated by
$\elp\elm\to\eta\rhoz$ (where the $\rhoz$ mixes with the $\omega$), while the non-resonant $\elp\elm\to\eta\pip\pim$ decay is much smaller. Due to the limited statistics, the non-resonant decay is not considered in the fitting formula. The process with a $\rhoz$ intermediate state violates G parity and is therefore a pure EM process, while that with an $\omega$ intermediate state can also proceed through the strong interaction.
%$\elp\elm\to\eta\rhoz\to\eta\pip\pim$ decays through pure EM interaction due to the G-parity violation,
%while that with \textcolor{black}{an} $\omega$ intermediate state $\elp\elm\to\eta\omega\to\eta\pip\pim$ partly through strong interaction.
Thus, the Born cross section for $\eta\pip\pim$ can be written as
\begin{eqnarray}
\label{eq_etapipi}
\sigma^{0}(W)=\left(\frac{\mathcal{A}}{W^2}\right)^2\frac{4\pi\alpha^{2}}{W^2}
\left|1+\frac{3W^2\sqrt{\Gamma_{ee}\Gamma_{\mu\mu}}\mathcal{C}_{1}e^{i\Phi_{\gamma,\rm cont}}(1+\mathcal{C}_{2}e^{i\Phi})}{\alpha M(W^2-M^{2}+iM\Gamma)}\right|^{2}, \notag
\end{eqnarray}
where {$\mathcal{C}_{1}$ represents the contribution from $\elp\elm\to\eta\rhoz$},
and $\mathcal{C}_{2}$ and $\Phi$ are the ratio and the relative phase between the processes
$\jpsi\to\eta\omega$ and $\jpsi\to\eta\rhoz$, respectively. The ratio between the branching fractions
$\frac{\mathcal{B}(\jpsi\to\eta\omega,\omega\to\pip\pim)}{\mathcal{B}(\jpsi\to\eta\rhoz,\rhoz\to\pip\pim)}$ is
$(0.138^{+0.025}_{-0.026})$ according to PDG~\cite{pdg}.
Thus, $\mathcal{C}_{2}$ is fixed to be $\sqrt{0.138}$ in the fit.
The branching fraction of $\jpsi\to\eta\pip\pim$ is determined by
$(\frac{\mathcal{A}}{W^2})^{2}|\mathcal{C}_{1}e^{i\Phi_{\gamma,\rm cont}}(1+\mathcal{C}_{2}e^{i\Phi})|^{2}\mathcal{B}(\jpsi\to\mup\mum)$.
Analogously, fixing the $\Gamma_{ee}$, $\Gamma_{\mup\mum}$ and $\Gamma$ as well as $M$ and $S_{E}$,
and assuming the relative phase $\Phi$ to be $0^{\circ}$ or $90^{\circ}$ for two extreme cases since it cannot be extracted from the fit,
the fit yields $\Phi_{\gamma,\rm cont}$ to be {$(-2\pm36)^{\circ}$ or $(-22\pm36)^{\circ}$} with the same goodness of fit.
The fit curve with $\Phi=0^{\circ}$ is shown in Fig.~\ref{fig_5pi} (c),
where the lineshape of $\elp\elm\to\eta\pip\pim$ is very similar to that of the process $\elp\elm\to\mup\mum$,
but different from that of $\elp\elm\to 5\pi$.
The branching fractions in the two cases are both calculated to be {$(3.78\pm0.66)\times10^{-4}$}.
\textcolor{black}{In alternative fits} with floating $M$ and $S_{E}$, changing $\Gamma_{ee}$, $\Gamma_{\mup\mum}$, $\Gamma$ and $\mathcal{C}_{2}$
by one standard deviation, or using the minimization method without \textcolor{black}{the factor $f$},
neither the phase nor the branching fraction changes.
Taking differences of various fit results as the systematic uncertainties, the phase $\Phi_{\gamma,\rm cont}$
is determined to be {$(-2\pm36)^{\circ}$ or $(-22\pm36)^{\circ}$}, and the $\mathcal{B}(\jpsi\to\eta\pip\pim)$ is {$(3.78\pm0.68)\times10^{-4}$}.

\section{SUMMARY} \label{summary}
For the first time, the relative phase between strong and EM amplitudes is measured directly from the $\jpsi$ lineshape.
Our result {$\Phi_{g,\rm EM}$ being $(84.9\pm3.6)^{\circ}$ or $(-84.7\pm3.1)^{\circ}$} from the $\elp\elm\to5\pi$ channel
confirms the orthogonality between the strong and EM amplitudes and supports the hypothesis of {a universal} phase in $\jpsi$ \textcolor{black}{decays}.
The relative phase between EM amplitudes from $\jpsi$ \textcolor{black}{decays} and from a virtual photon production in $\elp\elm$ interactions,
$\Phi_{\gamma,\rm cont}$ is determined from \textcolor{black}{the} $\mup\mum$ and $\eta\pip\pim$ processes, and
the results are consistent with the zero-phase ansatz.
Excluding the contribution from the continuum process and the interference, the branching fraction
of $\jpsi$ decays to $5\pi$ is measured to be
{$(4.73\pm0.44)\%$ or $(4.85\pm0.45)\%$}, which is consistent with the world average value \textcolor{black}{of} $(4.1\pm0.5)\%$~\cite{pdg}.
The branching fraction of $\jpsi\to\eta\pip\pim$ is {$(3.78\pm0.68)\times10^{-4}$},
which is \textcolor{black}{more accurate than} the existing world average value \textcolor{black}{of} $(4.0\pm1.7)\times10^{-4}$~\cite{pdg}.
All the uncertainties of \textcolor{black}{the} results are the statistical and systematic uncertainties added in quadrature.

\section{ACKNOWLEDGMENTS}

We thank the authors of \textcolor{black}{the} {\sc babayaga} and {\sc mcgpj} generators for their kind help and constructive discussion.
The BESIII collaboration thanks the staff of BEPCII and the IHEP
computing center for their strong support. This work is supported
in part by Ministry of Science and Technology of China on
the joined China-Italian project under Contract No. 2015DFG02380,
National Key Basic Research Program of China under Contract
No. 2015CB856700, 2015NSFC11375206;
National Natural Science Foundation of China (NSFC) under Contracts
Nos. 11235011, 11335008, 11425524, 11625523, 11635010;
the Chinese Academy of Sciences (CAS) Large-Scale Scientific Facility
Program; the CAS Center for Excellence in Particle Physics (CCEPP);
Joint Large-Scale Scientific Facility Funds of the NSFC and CAS under Contracts Nos. U1332201, U1532257, U1532258;
CAS under Contracts Nos. KJCX2-YW-N29, KJCX2-YW-N45;
CAS Key Research Program of Frontier Sciences under
Contracts Nos. QYZDJ-SSW-SLH003, QYZDJ-SSW-SLH040;
100 Talents Program of CAS; National 1000 Talents Program of China;
INPAC and Shanghai Key Laboratory for Particle Physics and Cosmology;
German Research Foundation DFG under Contracts Nos. Collaborative
Research Center CRC 1044, FOR 2359; Istituto Nazionale di Fisica Nucleare, Italy;
Koninklijke Nederlandse Akademie van Wetenschappen (KNAW) under Contract No. 530-4CDP03;
Ministry of Development of Turkey under Contract No. DPT2006K-120470;
National Natural Science Foundation of China (NSFC) under Contracts Nos. 11505034, 11575077;
National Science and Technology fund; The Swedish Research Council;
U. S. Department of Energy under Contracts Nos. DE-FG02-05ER41374,
DE-SC-0010118, DE-SC-0010504, DE-SC-0012069; University of Groningen (RuG) and the
Helmholtzzentrum fuer Schwerionenforschung GmbH (GSI), Darmstadt;
WCU Program of National Research Foundation of Korea under Contract No. R32-2008-000-10155-0.

%\begin{widetext}
%\end{widetext}

%\begin{widetext}

\vskip 1cm

\appendix{\bf Appendix: Analytical formula of radiation corrected cross section}
\vskip 0.5cm

The details of the deduction of the analytical formula of $\elp\elm\to\mup\mum$ {around $\jpsi$} has been published in Ref.~\cite{wangp,fengzhi,zhouxy}.
Here, only the {final formulas are presented.}
The continuum cross section after efficiency correction $f^{\rm EC}_{i}$ can be approximately written as:
%\begin{small}
\begin{flalign}
\begin{split}
\sigma^{\rm C}
=\frac{A}{W^2} \left[1+\frac{\beta}{2}(2\ln\Xf-\ln(1-\Xf)+\frac{3}{2}-\Xf)+ \frac{\alpha}{\pi}(\frac{\pi^2}{3}-\frac{1}{2}) \right]. \notag
\end{split}
\end{flalign}
%\end{small}
The resonance and interference part is:
%\begin{small}
\begin{flalign}
\begin{split}
\label{eq_RI}
\sigma^{\rm{R+I}}(W) &= C_{1}(1+\delta)\cdot\left[ a^{\beta-2}\varphi(\cos\zeta,\beta)+\beta(\frac{X^{\beta-2}_{f}}{\beta-2}+\frac{X^{\beta-3}_{f}}{\beta-3}R_{2}+\frac{X^{\beta-4}_{f}}{\beta-4}R_{3}) \right]~~~~~~~~~~~~~~~~~~~~~~~~~~~~~~~~~~~~\\ \notag
&+\left[ -\beta(1+\delta)C_{2}+(-\beta-\frac{\beta^{2}}{4})C_{1} \right]\cdot\left[ \frac{a^{\beta-1}}{1+\beta}\varphi(\cos\zeta,\beta+1)+\frac{X^{\beta-1}_{f}}{\beta-1} + \frac{X^{\beta-2}_{f}}{\beta-2}R_{2}+\frac{X^{\beta-3}_{f}}{\beta-3}R_{3} \right] \\ \notag
&+\left[ \frac{1}{2}\ln\frac{X^2_f+2aX_f\cos\zeta+a^2}{a^2}-{\rm ctg}\zeta(\textcolor{black}{\rm tg}^{-1}\frac{X_f+a\cos\zeta}{a\sin\zeta}-\frac{\pi}{2}+\zeta) \right]\cdot\left[(\beta+\frac{\beta^2}{4})\cdot C_{2}+(\frac{\beta}{2}-\frac{3}{8}\beta^2)\cdot C_{1}  \right]. \notag
\end{split}
\end{flalign}
%\end{small}

Here $a^2=(1-\frac{M^2}{W^2})^2+\frac{M^2\Gamma^2}{W^4}$, $\cos\zeta=\frac{1}{a}(\frac{M^2}{W^2}-1)$,
$\varphi(\cos\zeta,y)=\frac{\pi y\sin[\zeta(1-y)]}{\sin\zeta\sin\pi y}$, $R_{2}=\frac{2(W^2-M^2)}{W^2}=-2a\cos\zeta$, $R_{3}= a^2(4\cos^2\zeta-1)$,
$X_{f}=1-\frac{W^2_{\rm min}}{W^2}$. For $\elp\elm\to\mup\mum$, $C_{1}$ and $C_{2}$ are:
%\begin{small}
\begin{flalign}
\begin{split}
&C_{1} = \left\{8\pi\alpha^2 \frac{\sqrt{\Gamma_{ee}\Gamma_{\mu\mu}}}{M}\left[(W^2-M^2)\cos\Phi_{\gamma,\rm cont}+\sin\Phi_{\gamma,\rm cont} M\Gamma\right] + \frac{12\pi\Gamma_{ee}\Gamma_{\mu\mu}}{M^2}W^2  \right\}/W^4, \\ \notag
&C_{2} = \left[ \frac{8\pi\alpha\sqrt{\Gamma_{ee}\Gamma_{\mu\mu}}}{M}\cos\Phi_{\gamma,\rm cont} + \frac{12\pi\Gamma_{ee}\Gamma_{\mu\mu}}{M^2}  \right]/W^2. \\ \notag
\end{split}
\end{flalign}
%\end{small}

For $\elp\elm\to5\pi$, the analytical formula is very similar to $\elp\elm\to\mup\mum$, but the $C_{1}$ and $C_{2}$ are changed as:
\begin{eqnarray} \notag
\label{eq_5pi_C}
C_{1} &=& \{8\pi\alpha^2 \frac{\sqrt{\Gamma_{ee}\Gamma_{\mu\mu}}}{M}
\left[(W^2-M^2)(\mathcal{C}\cos\Phi_{g,\rm EM}+1)+\mathcal{C}\sin\Phi_{g,\rm EM} M\Gamma\right] \\\notag
&+& \frac{12\pi\Gamma_{ee}\Gamma_{\mu\mu}}{M^2}W^2(1+\mathcal{C}^{2}+2\mathcal{C}\cos\Phi_{g,\rm EM})  \}/W^4, \\ \notag
C_{2}&=& \left[ \frac{8\pi\alpha\sqrt{\Gamma_{ee}\Gamma_{\mu\mu}}}{M}(\mathcal{C}\cos\Phi_{g,\rm EM}+1)
- \frac{12\pi\Gamma_{ee}\Gamma_{\mu\mu}}{M^2}(1+\mathcal{C}^{2}+2\mathcal{C}\cos\Phi_{g,\rm EM})  \right]/W^2.
\end{eqnarray}

For $\eta\pip\pim$, the corresponding $C_{1}$ and $C_{2}$ are changed to
%\begin{small}

\begin{eqnarray}
\label{eq_etapipi2} \notag
C_{1} &=&\{8\pi\alpha^2 \frac{\sqrt{\Gamma_{ee}\Gamma_{\mu\mu}}}{M}
[(W^2-M^2)\mathcal{C}_{1}(\cos\Phi_{\gamma,\rm cont}+\mathcal{C}_{2}\cos\Phi_{\gamma,\rm cont}\cos\Phi-\mathcal{C}_{2}\sin\Phi_{\gamma,\rm cont}\sin\Phi) \\ \notag
&+& M\Gamma\mathcal{C}_{1}(\sin\Phi_{\gamma,\rm cont}+\mathcal{C}_{2}\sin\Phi_{\gamma,\rm cont}\cos\Phi+\mathcal{C}_{2}\cos\Phi_{\gamma,\rm cont}\sin\Phi)]\\ \notag
&+& \frac{12\pi\Gamma_{ee}\Gamma_{\mu\mu}}{M^2}W^2\mathcal{C}^{2}_{1}(1+\mathcal{C}^{2}_{2}+2\mathcal{C}_{2}\cos\Phi) \}/W^4, \\ \notag
C_{2} &=&[ \frac{8\pi\alpha\sqrt{\Gamma_{ee}\Gamma_{\mu\mu}}}{M}
\mathcal{C}_{1}(\cos\Phi_{\gamma,\rm cont}+\mathcal{C}_{2}\cos\Phi_{\gamma,\rm cont}\cos\Phi-\mathcal{C}_{2}\sin\Phi_{\gamma,\rm cont}\sin\Phi) \\ \notag
&+& \frac{12\pi\Gamma_{ee}\Gamma_{\mu\mu}}{M^2}\mathcal{C}^{2}_{1}(1+\mathcal{C}^{2}_{2}+2\mathcal{C}_{2}\cos\Phi)  ]/W^2. \notag
\end{eqnarray}

%\end{widetext}
%\end{multicols}
%\end{lstlisting}
\end{document}